\documentclass{aa}
\usepackage{amsmath}
\usepackage{cases}
\usepackage{txfonts}
\usepackage{graphicx}
\usepackage{natbib}
\usepackage{epstopdf}
\usepackage{nicefrac}
\usepackage{hyperref}

\newcommand{\Halpha}{$\mathrm{H}\alpha$ }
\newcommand{\SigmaPhot}{$\delta_{\mathrm{phot}}$ }
\bibpunct{(}{)}{;}{a}{}{,}

\begin{document}

\title{Extracting \Halpha flux from photometric data in the J-PLUS survey}

\author{G.~Vilella-Rojo\inst{1}
\and K.~Viironen\inst{1}
\and C.~L\'opez-Sanjuan\inst{1}
\and A.J.~Cenarro\inst{1}
\and J.~Varela\inst{1}
\and L.A. D\'iaz-Garc\'ia\inst{1}
\and D.~Crist\'obal-Hornillos\inst{1}
\and A.~Ederoclite\inst{1}
\and A.~Mar\'in-Franch\inst{1}
\and M.~Moles\inst{1}
}

\institute{Centro de Estudios de F\'{\i}sica del Cosmos de Arag\'on, Plaza San Juan 1, 44001 Teruel, Spain\\ \email{gvilella@cefca.es} \label{1}
}

\date{Submitted --, 2015}

\abstract
{}
{We present the main steps that will be taken to extract H$\alpha$ emission flux from Javalambre Photometric Local Universe Survey (J-PLUS) photometric data.}
{For galaxies with $z\lesssim0.015$, the H$\alpha$+[\ion{N}{ii}] emission is covered by the J-PLUS narrow-band filter $F660$. We explore three different methods to extract the \Halpha + [\ion{N}{ii}] flux from J-PLUS photometric data: a combination of a broad-band and a narrow-band filter ($r'$ and $F660$), two broad-band and a narrow-band filter ($r'$, $i'$ and $F660$), and an SED-fitting based method using eight photometric points. To test these methodologies, we simulated J-PLUS data from a sample of 7511 SDSS spectra with measured \Halpha flux. Based on the same sample, we derive two empirical relations to correct the derived H$\alpha$+[\ion{N}{ii}] flux from dust extinction and [\ion{N}{ii}] contamination.} 
{We find that the only unbiased method is the SED fitting based method. The combination of two filters underestimates the measurements of the \Halpha + [\ion{N}{ii}] flux by 22\%, while the three filters method are underestimated by 9\%. We study the error budget of the SED-fitting based method and find that, in addition to the photometric error, our measurements have a systematic uncertainty of 4.3\%. Several sources contribute to this uncertainty: the differences between our measurement procedure and that used to derive the spectroscopic values, the use of simple stellar populations as templates, and the intrinsic errors of the spectra, which were not taken into account. Apart from that, the empirical corrections for dust extinction and [\ion{N}{ii}] contamination add an extra uncertainty of 14\%.}
{Given the J-PLUS photometric system, the best methodology to extract \Halpha + [\ion{N}{ii}] flux is the SED-fitting based method. Using this method, we are able to recover reliable \Halpha fluxes for thousands of nearby galaxies in a robust and homogeneous way. Moreover, each stage of the process (emission line flux, dust extinction correction, and [\ion{N}{ii}] decontamination) can be decoupled and improved in the future. This method ensures reliable \Halpha measurements for many studies of galaxy evolution, from the local star formation rate density, to 2D studies in spatially well-resolved galaxies or the study of environmental effects, up to $\mathrm{m}_{r'}=21.8$ (AB; 3$\sigma$ detection of H$\alpha$+[\ion{N}{ii}] emission).}


\authorrunning{Vilella Rojo et al.}
\titlerunning{Robust estimation of H$\alpha$ flux with J-PLUS data}
\maketitle

\section{Introduction}\label{intro}
One of the most important processes driving the evolution of galaxies is the rate at which their gas is transformed into stars, namely the star formation rate (SFR). This parameter accounts for the amount of gas that is transformed into stars per unit time in a galaxy or a region of a galaxy. When the SFR is computed within a given cosmological volume, a SFR density $\left(\rho_{\mathrm{SFR}}\right)$ is obtained. 
\newline 
\indent The most direct indicator of the star-forming process is the ultraviolet (UV) radiation produced in the photosphere of young stars with intermediate to high masses ($\mathrm{M}\gtrsim3\mathrm{M_{\sun}}$). These stars emit more energy at short wavelengths ($\lambda<3000\AA$) than at longer wavelengths. As these stars are massive, their lifetimes are shorter than 300 Myr. Hence, UV traces star formation episodes of these timescales. The UV photons are, however, severely affected by dust extinction, which makes it difficult to translate UV flux into an SFR. Additionally, Earth's atmosphere blocks most of UV radiation, which makes  ground-based surveys in this wavelength range impossible.\newline 
\indent The UV radiation field of the most massive O stars ($\mathrm{M}\gtrsim20\mathrm{M_{\sun}}$) ionize the gas surrounding newborn stars. Hydrogen recombination leads to emission lines in different wavelength ranges, creating the Balmer series, among others. Thus, another widely used SFR indicator is the H$\alpha$ emission that is originated in the nebular cloud surrounding the star-forming region. This emission is prominent during the first $\sim10$ Myr of the star formation process, as the stars that can produce ionising photons die in these timescales. Hence, \Halpha emission probes more recent bursts than the UV. Moreover, this emission is less affected by dust than the UV continuum. Several calibrations exist to relate \Halpha flux with the SFR \citep[e.g. ][]{Kennicutt1998,Calzetti2013}.\newline
\indent Apart from these, there are other indirect SFR indicators. For example, the infrared (IR) thermal emission of dust heated by the UV field (see, for instance, \citealt{Calzetti2007} or  \citealt{PerezGonzalez2008}), or the forbidden [\ion{O}{ii}]$_{\lambda\lambda3727,\:3729}$ doublet \citep[][]{Kewley2004,Sobral2012}. The latter is useful for studies in the optical for objects at $z\gtrsim0.4$, when H$\alpha$ emission is shifted to the IR.\newline
\indent The SFR has been studied looking at \Halpha for more than 20 years, both in the local Universe and at higher redshifts. For example, the study of \cite{Gallego1995} computed the SFR density in the local Universe with the UCM Survey data \citep{Zamorano1994,Gallego1995}. The work by \cite{Ly2007} uses narrow-band photometry to derive the luminosity function (LF) and the SFR at $0.07\leq z\leq1.47$ using information of H$\alpha$ and [\ion{O}{ii}]. Finally, the High Redshift Emission Line Survey  \citep[HiZELS, ][]{Geach2008} also studied \Halpha emission at $z=2.23,1.47,0.84$ and 0.40, using narrow-band photometry to estimate the SFR evolution of the last $\sim11$ Gyr of the Universe \citep{Sobral2013}.\newline
\indent With the Javalambre Photometric Local Universe Survey (J-PLUS\footnote{http://www.j-plus.es}; Cenarro et al. in prep) we aim to derive the SFR of the local Universe measuring the \Halpha emission of nearby galaxies ($z\lesssim0.015$). The design and survey strategy of J~-~PLUS allows us to probe the faint end of the LF($\mathrm{H}\alpha$). It is expected to reach $\sim10^{38}$ erg $\cdot$ s$^{-1}$, which is 2.5 orders of magnitude deeper than the UCM Survey and the KPNO International Spectroscopic Survey \citep[KISS, ][]{Salzer2000,Salzer2001}. In addition, thanks to the large surveyed volume, the bright end of the luminosity function is also constrained.\newline

The goal of this paper is to find the best method to extract H$\alpha$ fluxes from the J-PLUS photometric data. We present the methodology to analyse J-PLUS spectral energy distributions (SEDs), and the estimation of the errors that arise with this methodology. The paper is organized as follows: Section~\ref{J-PLUS} summarizes the J-PLUS photometric system and the main characteristics of the $\mathrm{J-PLUS}$ survey. Section~\ref{metodologias} presents the different methodologies that can be used to recover emission line fluxes from photometric data with a combination of broad-band and narrow-band filters, and in concrete the emission of H$\alpha$ + [\ion{N}{ii}]. In Section~\ref{simulaciones} we test these methodologies with simulated J-PLUS photo-SEDs, based on a sample of 7511 Sloan Digital Sky Survey (SDSS) spectra of star-forming galaxies with measured H$\alpha$ fluxes. In Section~\ref{simuls_diferentes_mags} we study the errors involved in the methodology. Dust and [\ion{N}{ii}] corrections are treated 
in Section~\ref{Polvo y nitrógenos}. Finally, in Section~\ref{summary} we give the most relevant results and conclusions of the work. Throughout this work, magnitudes are given in the AB system (\citealt{Oke83}).

\section{The J-PLUS Survey}\label{J-PLUS}
The J-PLUS survey is an auxiliary survey of the Javalambre-PAU Astrophysical Survey (J-PAS; \citealt{BenitezJPAS}). J-PLUS will be carried out at the Observatorio Astrof\'isico de Javalambre (OAJ; \citealt{CenarroOAJ}) using the Javalambre Auxiliary Survey Telescope (JAST/T80) and T80Cam (\citealt{ToniT80Cam}). This is a wide-field camera (2.1 $\mathrm{deg}^2$ per exposure) installed at the Cassegrain focus. It is equipped with a 9.2k-by-9.2k CCD with a pixel scale of ~0.55\arcsec pixel\textsuperscript{-1}.

\begin{figure}
  
  \centering
    \includegraphics[width=0.5\textwidth]{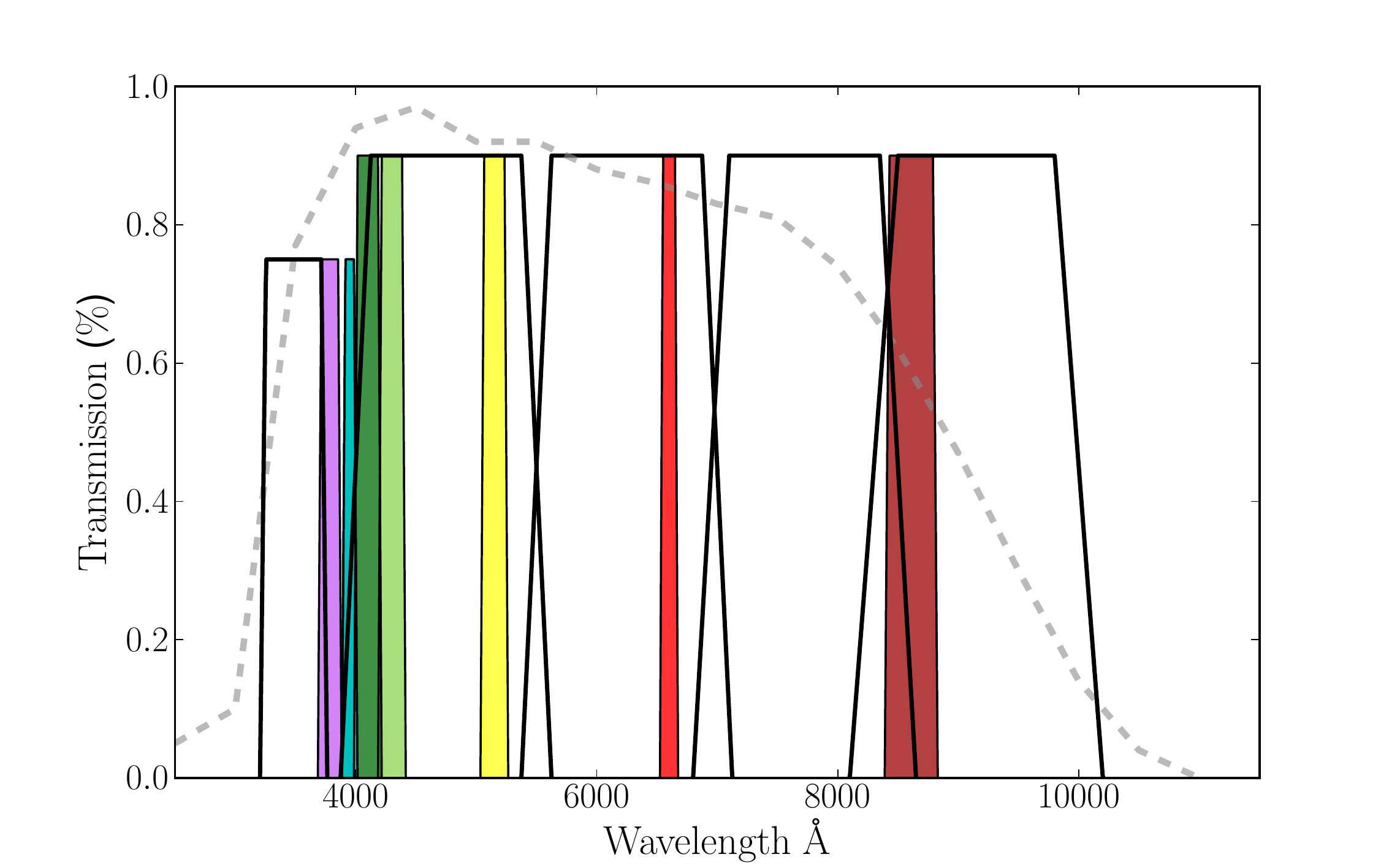}
    \caption{J-PLUS photometric system. Solid lines are the theoretical transmission curves of the broad-band J-PLUS filters. Coloured areas denote the theoretical transmission curves of the narrow-band J-PLUS filters. The dashed line represents the quantum efficiency of the T80Cam.}
    \label{Filtros}
    
\end{figure}

Expected to start in 2015, J-PLUS will cover an area of $\sim8500\: \deg^{2}$ of the northern sky, and has been particularly designed to carry out the photometric calibration of J-PAS using a specific set of 12 filters: 5 broad-band, and 7 narrow-band filters located at key stellar spectral features. Fig.~\ref{Filtros} shows the transmission curves of the filters. This filter set allows us to retrieve accurate SEDs for millions of calibration stars, which will be used to transport the photometric calibration from J-PLUS to J-PAS \citep{Gruel2012}.\newline \indent In addition to the calibration goals, J-PLUS has been designed to acquire the H$\alpha$ flux of the galaxies in the nearby Universe ($z\leq0.015$) using one of the narrow-band filters ($F660$) up to a detection magnitude $\sim22.5$ in that band (see Table~\ref{J-PLUS_tabla} for details of the J-PLUS photometric system). Moreover, the large field of view and efficiency of JAST/T80 and T80Cam make J-PLUS a powerful survey for other galaxy evolution 
studies,
 particularly those that require spatially well-resolved galaxies.

\begin{table*}
\caption{J-PLUS photometric system properties.}
\label{J-PLUS_tabla}
\begin{center}
\begin{tabular}{lccccc}
\hline\hline\noalign{\smallskip}

Filter      &    Central             &    FWHM    &    mag (AB)   & mag (AB)  & Comments     \\
name       &      wavelength (nm) &   (nm)  &    S/N=3      & S/N=50    &    		  \\
\noalign{\smallskip}
\hline
\noalign{\smallskip}
	$u'_J$ &  348.5		& 50.8  & 23.00  & 19.69 & In common with J-PAS  			\\
	$F378$   &  378.5		& 16.8 	& 21.58  & 18.01 & [\ion{O}{ii}]; in common with J-PAS			\\
	$F395$   &  395.0		& 10.0 	& 21.60  & 18.01 & Ca H+K					\\
	$F410$   &  410.0		& 20.0 	& 21.62  & 18.03 & H$\delta$					\\
	$F430$   &  430.0		& 20.0 	& 21.65  & 18.01 & G-band					\\
	$g'$     &  480.3		& 140.9	& 23.23  & 19.92 & SDSS 					\\
	$F515$   &  515.0		& 20.0 	& 21.67  & 18.01 & Mg						\\
	$r'$     &  625.4		& 138.8	& 23.26  & 19.99 & SDSS						\\
	$F660$   &  660.0		& 13.8 	& 22.64  & 19.11 &  H$\alpha$+[\ion{N}{ii}]; in common with J-PAS	\\
	$i'$     &  766.8		& 153.5	& 22.31  & 18.88 & SDSS						\\
	$F861$   &  861.0		& 40.0 	& 21.48  & 18.01 & Ca Triplet  				\\
	$z'$     &  911.4		& 140.9	& 21.51  & 18.19 & SDSS  					\\

\noalign{\smallskip}
\hline
\end{tabular}
\end{center}
\end{table*}

\section{Methodologies}\label{metodologias}
In this section, we present a collection of methods that can be used to obtain the flux of H$\alpha$ + [\ion{N}{ii}] $\lambda\lambda6548$, $6584$ with J-PLUS photometric data. The width ($\sim150\AA$) and central wavelength ($6600\:\AA$ )  of $F660$ filter ensures that we enclose the three lines inside the filter at $z\lesssim 0.015$. We present the basic assumptions of each method and the most relevant equations in the following sections.

\subsection{Two filters (\emph{2F}) method}\label{metodologias 2 filtros}
The simplest method that can be used is a combination of two filters: one to trace the continuum, and another to contrast the emission line. This can be achieved with either two adjacent or overlapping  filters. Given the J-PLUS filter system, we test the case that involves a broad filter to trace the continuum (\emph{r'}), and a narrow filter placed at the wavelength range of the line of interest ($F660$). This methodology is widely used in many photometric studies (e.g. \citealt{Ly2007}, \citealt{Takahashi2007}, \citealt{Villar2008}, \citealt{Koyama2014}, \citealt{An2014}); see also the studies of the \Halpha galaxy survey \citep{James2004} for a combination of non-overlapping narrow-band filters. \newline

\indent In this case, and taking into account that \emph{r'} contains  the flux of the H$\alpha$ + [\ion{N}{ii}] lines,  the flux of these emissions can be recovered following the standard recipe \citep[see][for a detailed description]{Pascual2007}:

\begin{equation}\label{ecuacion2filtros}
F_{\mathrm{H}\alpha+\left[\mathrm{NII}\right]}=\Delta_{F660}\frac{\left(\overline{F}_{F660}-\overline{F}_{r'}\right)}{1-\frac{\Delta_{F660}}{\Delta_{r'}}}\:, 
\end{equation}

\noindent where $\overline{F}_{F660}$ and $\overline{F}_{r'}$  are the observed average fluxes inside the $F660$ and $r'$ filters, and $\Delta_{x}$ is defined for any passband \emph{x} at any wavelength of interest $\lambda_{\mathrm{s}}$ as

\begin{equation}
\label{Definicion_delta}
 \Delta _{\mathrm{x}}\equiv \frac{\int P_{\mathrm{x}}\left(\lambda\right)\lambda \mathrm{d}\lambda}{P_{\mathrm{x}}  \left(   \lambda=\lambda_{\mathrm{s}}\right)  \lambda_{\mathrm{s}}}\:,
\end{equation}

\noindent where $P_{x}$ is the transmission of the passband \emph{x} as a function of wavelength. In our case, $\lambda_{\mathrm{s}}=\lambda_{\mathrm{H}\alpha}$ at $z=0$, i.e. $\lambda_{\mathrm{s}}=6562.8\:\AA$. For J-PLUS $F660$ and \emph{r'} filters, we found $\Delta _{F660}=125.3\:\AA$ and $\Delta _{r'}=1419\:\AA$. The strongest assumption of this approximation is a flat continuum inside the \emph{r'} filter.

\subsection{Three filters (\emph{3F}) method}\label{metodologias 3 filtros}
To solve the problem of the flat continuum assumption of the 2F method, we can use two filters to trace a linear continuum, and a narrow-band filter to contrast the line. There are different configurations for this method: in the work by \cite{Kennicutt1983} no overlapping between the three filters occurs, while the case of the narrow-band filter overlapping two broad filters is studied in \cite{Pascual2007}.\newline
\indent Because of the filter configuration of J-PLUS, we test the case in which the continuum is inferred using \emph{r'} and \emph{i'}, while the emission is inside $F660$. As in the 2F method, the $r'$ flux also contains the lines' flux, and it is risen from the stellar continuum. Because of this, we use an analytic formula to remove the H$\alpha$ + [\ion{N}{ii}] contribution inside the $r'$ filter, i.e. 

\begin{equation}\label{Ecflujo3filtro}
 F_{\mathrm{H}\alpha+\left[\mathrm{NII}\right]}=\frac{\left(\overline{F}_{r'}-\overline{F}_{i'}\right)-\left(\frac{\alpha_{r'}-\alpha_{i'}}
 {\alpha_{F660}-\alpha_{i'}}\right)\left(\overline{F}_{F660}-\overline{F}_{i'}\right)}{\beta_{F660}\left(\frac{\alpha_{i'}-\alpha_{r'}}{\alpha_{F660}-\alpha_{i'}}\right)+\beta_{r'}}\:,
\end{equation}

\noindent where $\overline{F}$ are the observed average fluxes, and $\alpha$ and $\beta$ are defined as

\begin{equation}
\alpha_{x}\equiv\frac{\int \lambda^{2}P_{x}\left(\lambda\right)\mathrm{d}\lambda}{\int P_{x}\left(\lambda\right)\lambda \mathrm{d}\lambda}\qquad,\qquad\beta_{x}\equiv\frac{\lambda_{s}\mathrm{·}P_{x}\left(\lambda=\lambda_{s}\right)}{\int P_{x}\left(\lambda\right)\lambda \mathrm{d}\lambda},
\end{equation}

\noindent where in our particular case, $\lambda_{s}=\lambda_{\mathrm{H}\alpha}=6562.8\:\AA$. These equations are detailed in the Appendix~\ref{Apendice 3 filtros}.

\subsection{SED-fitting method}\label{metodologias SED}
This method benefits from all the J-PLUS filters to infer the emission flux after fitting the stellar continuum of the galaxy. To do that, we compare simulated J-PLUS observations with a set of template models, and for each pair observation-template, we compute the value of the $\chi^{2}$ function, defined as

\begin{equation}
\chi_{j}^{2}=\sum_{x}\left(\frac{\overline{F}_{x}-k_{j}\overline{T}_{x}^{j}}{\delta_{x}}\right)^{2},\label{eq:formula ch}
\end{equation}

\noindent where $\overline{F}_{x}$ is the observed flux for each of the \emph{x} filter with its error $\delta_{x}$, $\overline{T}_{x}^{j}$ is the flux inside the \emph{x} filter of the $j$ template, and $k$ scales the templates at the magnitude of the galaxy that we are fitting. We estimate this scaling parameter for each \emph{j} template by minimizing Eq.~\ref{eq:formula ch}, i.e.

\begin{equation}
k_{j}=\frac{\sum_{x}\left(\frac{ \overline{F}_{x}\overline{T}_{x}^{j}}{\delta_{x}^{2}}\right)}{\sum_{x}\left(\frac{\left(\overline{T}_{x}^{j}\right)^{2}}{\delta_{x}^{2}}\right)}\label{eq:formula k}\:.
\end{equation}

Template models are simple stellar populations (SSPs) taken from \citet[][BC03 hereafter]{BC03}, with 40 ages from 1 Myr to 13.75 Gyr in logarithmic bins, a \cite{Salpeter55} initial mass function, and six metallicities ranging from $0.2\mathrm{Z}_{\odot}$ to $2.5\mathrm{Z}_{\odot}$. We extinguish these models with a \cite{Calzetti2000} law from $E(B-V)=0$ to $E(B-V)=1$ in steps of $0.05$. In the end we have 4200 templates at rest-frame. These are convolved with the J-PLUS photometric system properties using \texttt{PySynphot}\footnote{A \texttt{Python} programming language adaptation of the widely used Synphot, developed by the Space Telescope Science Institute (STSCI)}. The template fluxes are then

\begin{equation}
  \label{convolucionnn}
  \overline{T}_{x}^{j}  =\frac{ \int T_{\lambda}P_{x}\left(\lambda\right)\lambda \mathrm{d}\lambda }{\int P_{x}\left(\lambda\right)\lambda \mathrm{d}\lambda  }\:.
\end{equation}

The $\chi^{2}$ fitting is carried out without taking the flux of the $F660$ filter into account, which contains the H$\alpha$ + [\ion{N}{ii}] emission fluxes. To derive the emission flux, we approximate it as a single line described as a Dirac's delta function. This is called ``the infinite thin line approximation'' \citep[see, for instance][]{Pascual2007}. If we split the total $F660$ flux into two components (continuum and emission) and introduce it into Eq.~\ref{convolucionnn}, we get

\begin{equation}
\label{flujo dentro de f660}
 \overline{F}_{F660}=\frac{\int\left(F_{\mathrm{H}\alpha+\mathrm{[\ion{N}{ii}]}}+F_{\mathrm{cont}}\right)P_{F660}\left(\lambda\right)\lambda \mathrm{d}\lambda}{\int P_{F660}\left(\lambda\right)\lambda \mathrm{d}\lambda}\:,
\end{equation}

\noindent where $F_{\mathrm{H}\alpha+\mathrm{[\ion{N}{ii}]}}$ is the emission flux, and $F_{\mathrm{cont}}$ the continuum. If we assume that $F_{\mathrm{H}\alpha+\mathrm{[\ion{N}{ii}]}}\equiv F_{\mathrm{H}\alpha+\mathrm{[\ion{N}{ii}]}}\delta\left(\lambda-\lambda_{\mathrm{H}\alpha}\right) $, and split Eq.~\ref{flujo dentro de f660} in two integrals, we get

\begin{equation}
 \label{terminos separados}
 \overline{F}_{F660} = \overline{F}_{\mathrm{F660,\:cont}} + \frac{1}{\Delta_{F660}}{F}_{\mathrm{H}\alpha+\mathrm{[\ion{N}{ii}]}}\:.
\end{equation}

Because the $F660$ filter overlaps with \emph{r'}, the H$\alpha$ + [\ion{N}{ii}] fluxes are also inside the $r'$ filter, which leads to

\begin{equation}
 \label{flujo en r}
 \overline{F}_{r'} = \overline{F}_{r',\: \mathrm{cont}} + \frac{1}{\Delta_{r'}}{F}_{\mathrm{\mathrm{H}\alpha+\mathrm{[\ion{N}{ii}]}}}\:.
\end{equation}

To remove the \Halpha+[\ion{N}{ii}] contribution from the $r'$ filter, and obtain a more reliable \emph{r'} continuum, using Eq.~\ref{flujo en r} we decontaminate the \emph{r'} flux by subtracting the ${F}_{\mathrm{\mathrm{H}\alpha+\mathrm{[\ion{N}{ii}]}}}$ flux that is inferred with the 3F method.

\begin{figure}
  \centering
  \includegraphics[width=0.5\textwidth]{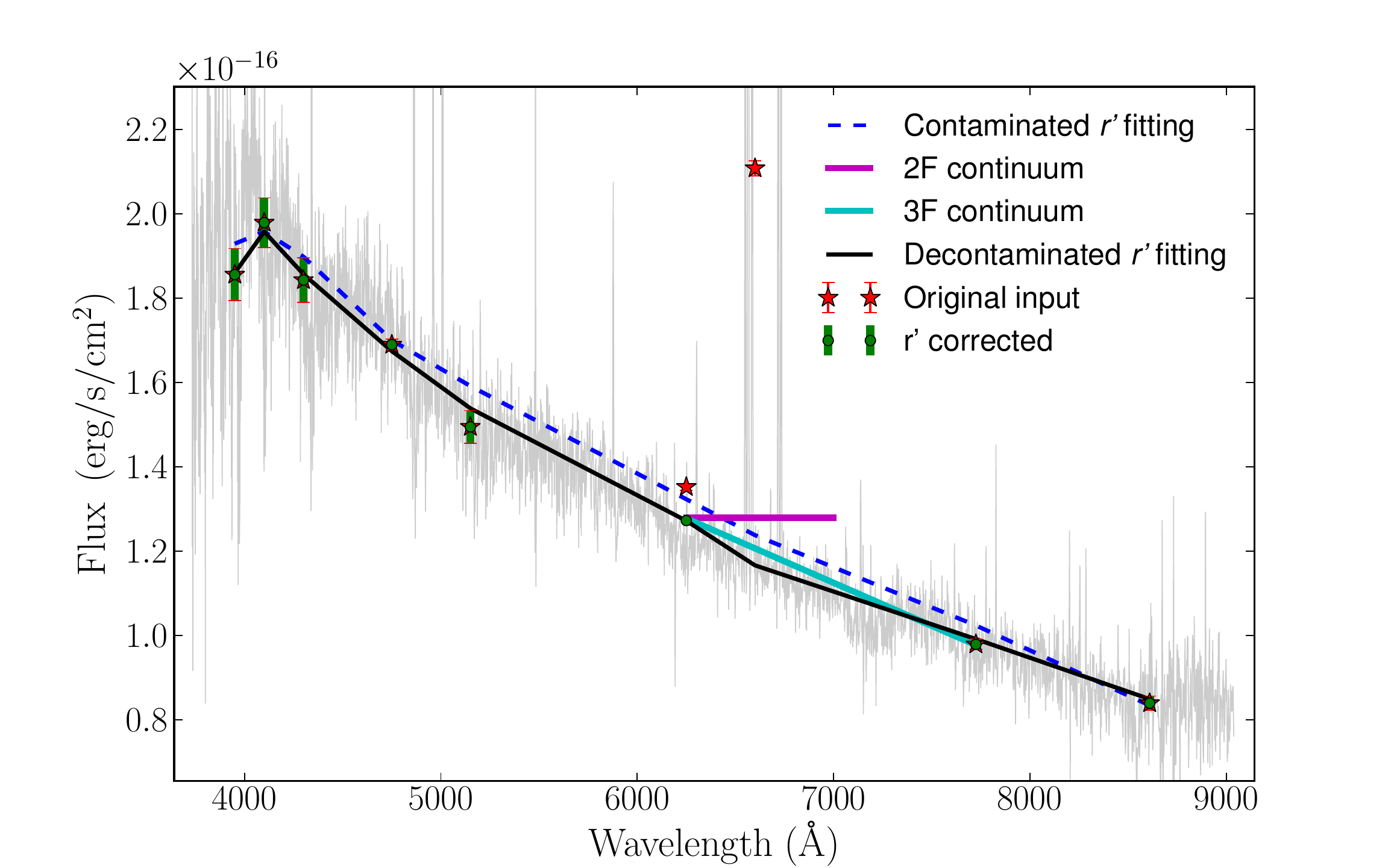}
  \caption{Example of a SED fitting plotted over the SDSS spectra. Red asterisks show the original input (note the risen \emph{r'} dot). Green dots denote the photometric points after \emph{r'} decontamination. These points overlap the red ones except for the $r'$, because the \emph{r'} correction does not take them into account. Magenta solid line indicates the \Halpha continuum using the 2F method. Cyan solid line shows the \Halpha continuum using the 3F method. Dashed blue line denotes the best-fitting template if no $r'$ decontamination is done. Black solid line is the best-fitting template after \emph{r'} decontamination.}
  \label{ejemploFit}
\end{figure}

We present an example of an SED fitting in Fig.~\ref{ejemploFit}, and compare the continuum that results from the equations of the 2F and 3F methodologies.

\subsection{Measurements and error estimation}\label{Mediana y sigmaphot}
To estimate the final \Halpha + [\ion{N}{ii}] flux and its error for each methodology, we perform 500 Monte Carlo runs. To do that, we perturb each passband \emph{x} flux within a Gaussian distribution with $\mu=\overline{F}_{x}$ and $\sigma=\delta_{x}$ and apply each methodology.  We keep a record of the inferred H$\alpha$ + [\ion{N}{ii}] flux and perturb the original data again. In the end, we have an array of 500 H$\alpha$ + [\ion{N}{ii}] flux measurements for each method, noted as $\mathbf{F}_{\mathrm{H}\alpha + \mathrm{[\ion{N}{ii}]}}$. Our final measurement for the H$\alpha$ + [\ion{N}{ii}] flux is the median of this array,
\begin{equation}
 \langle F_{\mathrm{H}\alpha + \mathrm{[\ion{N}{ii}]}}\rangle = \mathrm{median}\left( \mathbf{F}_{\mathrm{H}\alpha + \mathrm{[\ion{N}{ii}]}}\right),
\end{equation}
while the photometric error $\delta_{\mathrm{phot}}$ associated with this measurement is the median absolute deviation (MAD) of this array,

\begin{equation}
 \delta_{\mathrm{phot}}=1.48\times \mathrm{median}\left( \, |\,\mathbf{F}_{\mathrm{H}\alpha + \mathrm{[\ion{N}{ii}]}} - \langle F_{\mathrm{H}\alpha + \mathrm{[\ion{N}{ii}]}}\rangle\,|\, \right).
\end{equation}

\section{Testing the methodologies}\label{simulaciones}
\subsection{Data sample}
\begin{figure}
  \centering
  \includegraphics[width=0.5\textwidth]{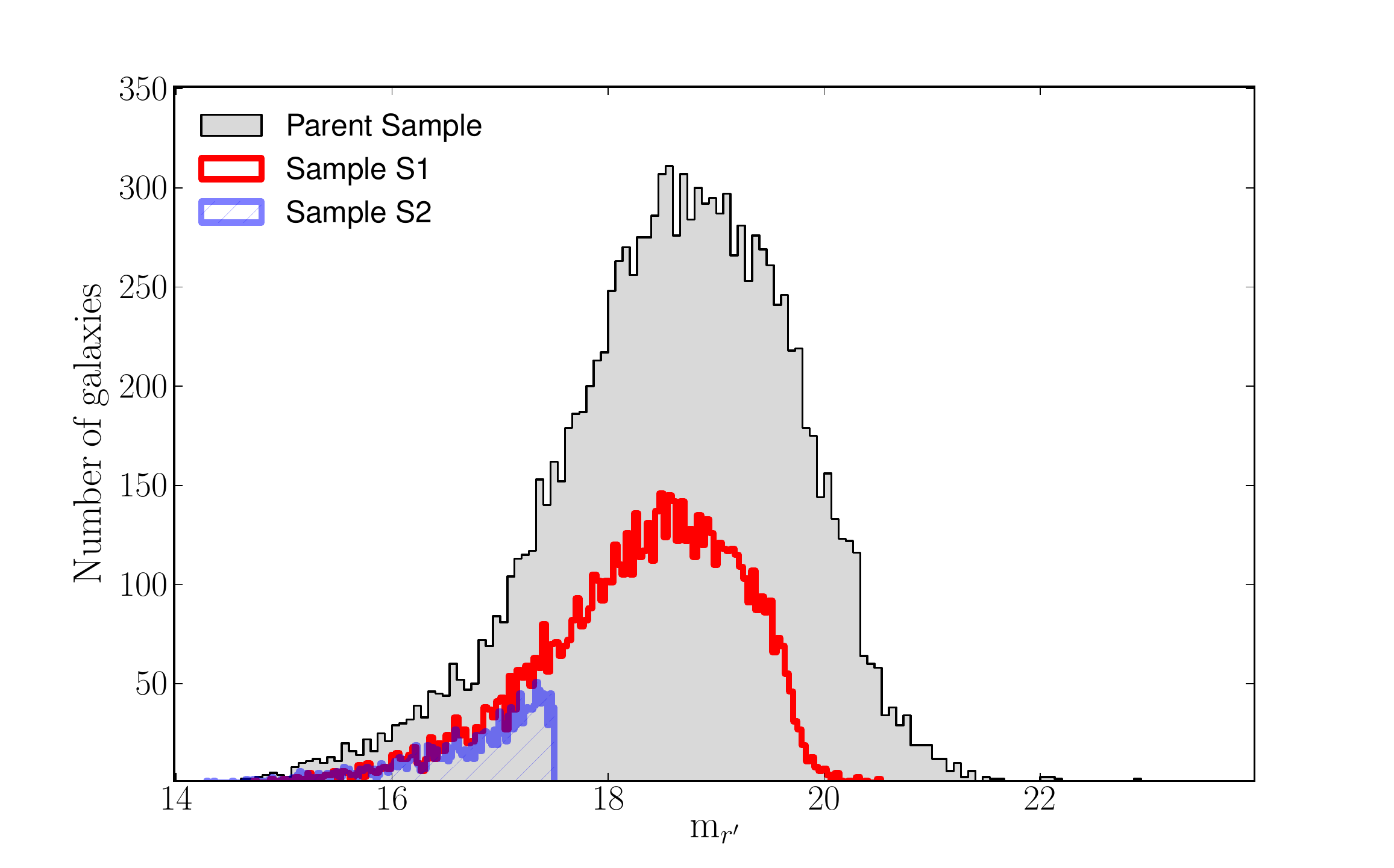}
  \includegraphics[width=0.5\textwidth]{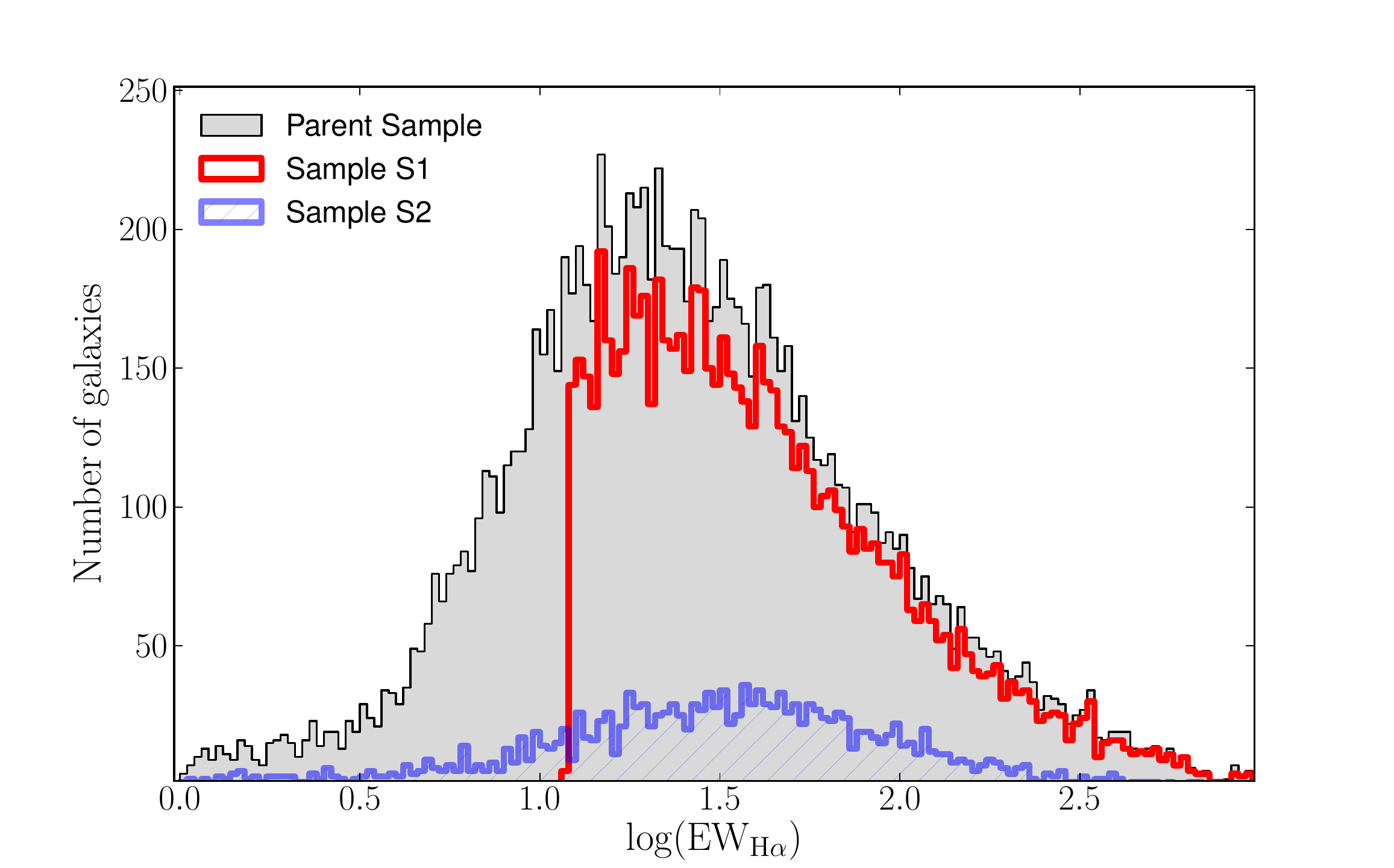}
  \caption{SDSS star-forming sources with $z<0.02$ (grey areas); \textbf{Top panel}: red line shows the sample S1 distributed as a function of $\mathrm{m}_{r'}$. Dashed area denotes the sample S2 (see Sect.~\ref{SampleS2}) distributed as a function of $\mathrm{m}_{r'}$. \textbf{Bottom panel}: red line shows sample S1 distributed as a function of $\log\mathrm{EW}_{\mathrm{H}\alpha}$. Dashed area illustrates the sample S2 (see Sect.~\ref{SampleS2}) distributed as a function of $\log\mathrm{EW}_{\mathrm{H}\alpha}$.   }
  \label{muestra_ew}
\end{figure}

To test each methodology, we use a set of SDSS spectra with emission lines measured by the Portsmouth Group \citep{Thomas2013} in the 10th Data Release (DR10). We excluded Barion Oscillation Spectroscopic Survey \citep[BOSS, ][]{Dawson2013} galaxies, as its targets are luminous red galaxies at $z\geq0.2$. We selected all the objects that were classified as star forming by the Portsmouth group, according to a BPT diagram criteria \citep{Baldwin1981}. Thus, we do not expect a significant AGN contamination in the sample. We applied a spectroscopic redshift cut at $z<0.02$. This redshift cut is chosen to probe a volume big enough to have a large number of galaxies with similar properties as the expected J-PLUS sources. After applying these criteria, we are left with $\sim$12000 spectra.
\newline
\indent From this sample, we retain only the spectra with an H$\alpha$ equivalent width ($\mathrm{EW}_{\mathrm{H}\alpha}$) $12\:\AA\geq \mathrm{EW}_{\mathrm{H}\alpha}$. This cut was done assuming that J-PLUS cannot resolve, with a precision of $3\sigma$, $\mathrm{EW}_{F660}\leq 12\:\AA$ because of the errors in the determination of magnitude of the zero point. From \cite{Pascual2007} we know that

\begin{equation}
 \mathrm{EW}=\Delta_{F660}\left(Q-1\right) \frac{Q-1}{1-Q\epsilon}\:,
\end{equation}

\noindent where $\epsilon \equiv \nicefrac{\Delta_{F660}}{\Delta_{r'}}$, and $2.5\log Q=\mathrm{m}_{r'}-\mathrm{m}_{F660}$. Assuming a systematic error in the determination of the zero-point magnitude of $\delta \mathrm{m}\approx0.02$ in \emph{r'} and $F660$, there is a limiting difference in magnitudes of $\delta \mathrm{m}\sim 0.03$, which we cannot resolve with enough confidence, and which leads to a minimum EW of detection that is $\mathrm{EW}_{F660}\simeq12\:\AA$. From the SDSS data we cannot know the observed $\mathrm{EW}_{F660}$, so we take $\mathrm{EW}_{\mathrm{H}\alpha}=12\:\AA$ as lower limit. Finally, a last cut in the median signal-to-noise ratio (S/N) of each spectrum was done. Only spectra with average $\mathrm{S/N}\geq5$ are kept. This limit is chosen to guarantee that spectroscopic flux errors are not important.\newline 
\indent After applying these criteria, the sample contains 7511 spectra. We refer to this sample as S1. Figure~\ref{muestra_ew} presents the distribution of the S1 as a function of the $\mathrm{EW}_{F660}$ and $\mathrm{m}_{r'}$. As, for now, the redshift is not taken into account in our analysis, we shifted all the spectra of S1 to $z=0$. After this, we convolved them with the J-PLUS filters that are in the wavelength range of the SDSS spectra to obtain $\overline{F}_{x}$. Because these spectra have a shorter wavelength range than J-PLUS filters, we lose the information of $u'_J$, $F378$, and $z'$ bands. The convolution retrieves the apparent magnitudes and mean fluxes of each passband. We stress that these apparent magnitudes are computed from the flux enclosed inside the fibre, and are not representative of the whole galaxy.  With the apparent magnitudes, we compute the expected S/N using the J-PLUS exposure time calculator\footnote{www.cefca.es/jplusetc}. This tool provides the estimated S/N given an 
apparent magnitude, an exposure time (that can be divided into several exposures) and different sky conditions. The assumed conditions to compute the S/N values were a grey night, a seeing of $0.9$ arcsec, a photometric aperture of $1.8$ arcsec, and an airmass of $1.2$. With this, the flux error in each passband \emph{x} is $\delta_{x}=\nicefrac{\overline{F}_{x}}{\mathrm{S/N}_x}$.
\newline 
\indent We apply each of the methodologies explained in Sect.~\ref{metodologias}, to the J-PLUS photo-spectra of S1 and study the performance of each methodology below.

\subsection{J-PLUS \emph{vs.} SDSS}\label{Simulations-results}
At this point we test the precision of each method described in Sect.~\ref{metodologias} to recover the flux of H$\alpha$ + [\ion{N}{ii}]. To this aim, we compare our inferred fluxes with those provided by the Portsmouth Group. As their values are dust corrected, but the SDSS spectra are not, we add dust to the SDSS measurements following the \cite{Calzetti2000} extinction law with $R_V~=~4.05$, which the Portsmouth group applied, and the values of $E(B-V)$ provided by them.\newline 
\indent We apply each methodology to the spectra of S1, and compare the recovered flux of \Halpha + [\ion{N}{ii}] lines with the spectroscopic measurements,

\begin{equation}\label{R}
  \mathrm{R}=  \frac{  \langle F_{\mathrm{H}\alpha + \mathrm{[\ion{N}{ii}]}}\rangle  }{F_{\mathrm{H}\alpha+\mathrm{[\ion{N}{ii}]}}^{\mathrm{SDSS}}},
\end{equation}

\noindent where, in this case, $F_{\mathrm{H}\alpha+\mathrm{[\ion{N}{ii}]}}^{\mathrm{SDSS}}$ is

\begin{equation}\label{polvoanadido}
 F_{\mathrm{H}\alpha+\mathrm{[\ion{N}{ii}]}}^{\mathrm{SDSS}}=\left[ F_{\mathrm{H}\alpha}+F_{\mathrm{[\ion{N}{ii}]},\,\lambda6548}+F_{\mathrm{[\ion{N}{ii}]},\,\lambda6583} \right]\times 10^{-1.33\cdot E(B-V)}
\end{equation}
because we are adding dust attenuation to its measurements. The resulting distribution of ratios R is fitted to a Gaussian in all the cases.

\subsubsection{2F and 3F methods}\label{Simulations - results 2,3 Filtros}
\begin{figure}
  \centering
  \includegraphics[width=0.5\textwidth]{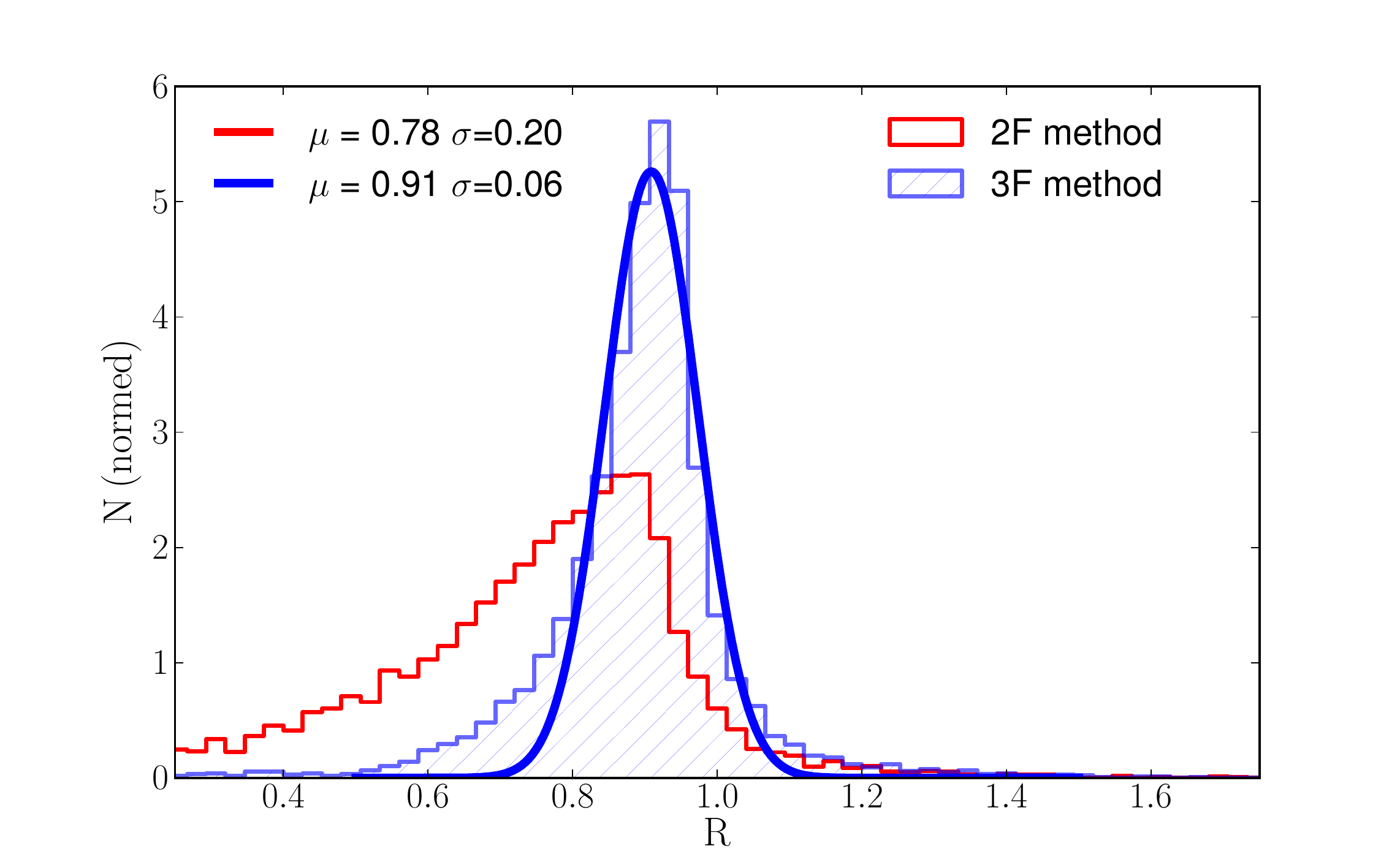}
  \caption{Normalized distribution of the ratios between the recovered and the spectroscopically measured H$\alpha$ + [\ion{N}{ii}] flux for S1 spectra, with the 2F (empty red histogram) and 3F (dashed blue histogram) methods. Blue curve shows the Gaussian fitting to the distribution of R in the 3F method. The best fitting values are labelled in the panel.}
  \label{gaussianas2y3Filtros}
\end{figure}

We show the resulting distribution of R when we analyse the spectra applying the 2F and 3F methodologies in Fig.~\ref{gaussianas2y3Filtros}. The 2F method is biased, mostly because we assume a flat continuum that is given by the broad filter. We lose the information about the colour of the galaxy, i.e. the true shape of the continuum at the wavelength range of the emission line. The distribution of results is very asymmetric, and cannot be fitted to a Gaussian distribution. The median of this distribution is $\mu=0.78$, while the dispersion is given by $\sigma_{\mathrm{2F}}=0.5\times(P_{84}-P_{16})=0.20$, where $P_{84}$ and $P_{16}$ are the $84^{\mathrm{th}}$ and $16^{\mathrm{th}}$ percentiles, respectively.
\newline
\indent To cope with this bias, \cite{Sobral2009,Sobral2012}, and subsequent works from the HiZELS survey, and also the work by \cite{Ly2011}, introduce a correction based on the broad-band to narrow-band colour. This correction is applied to compensate the flat continuum assumption and the difference between the central wavelength of the broad-band filter and the narrow-band central wavelength. This correction would be a step between the 2F method and the 3F method. However, we prefer the 3F method, as its correction is analytic.\newline
\indent With the 3F method, the distribution of results becomes almost Gaussian, but a bias of $\sim9\%$ still persists, with $\mu=0.91$ and a dispersion $\sigma_{\mathrm{3F}}=0.06$ from the Gaussian fit. In this case, we are more sensitive to the true shape of the continuum in the wavelength range of the line, but we still have to assume it is linear, which is a poor approximation because of the H$\alpha$ absorption. 

\subsubsection{SED fitting}\label{Simulations - results SED}
\begin{figure}
  
  \centering
  \includegraphics[width=0.5\textwidth]{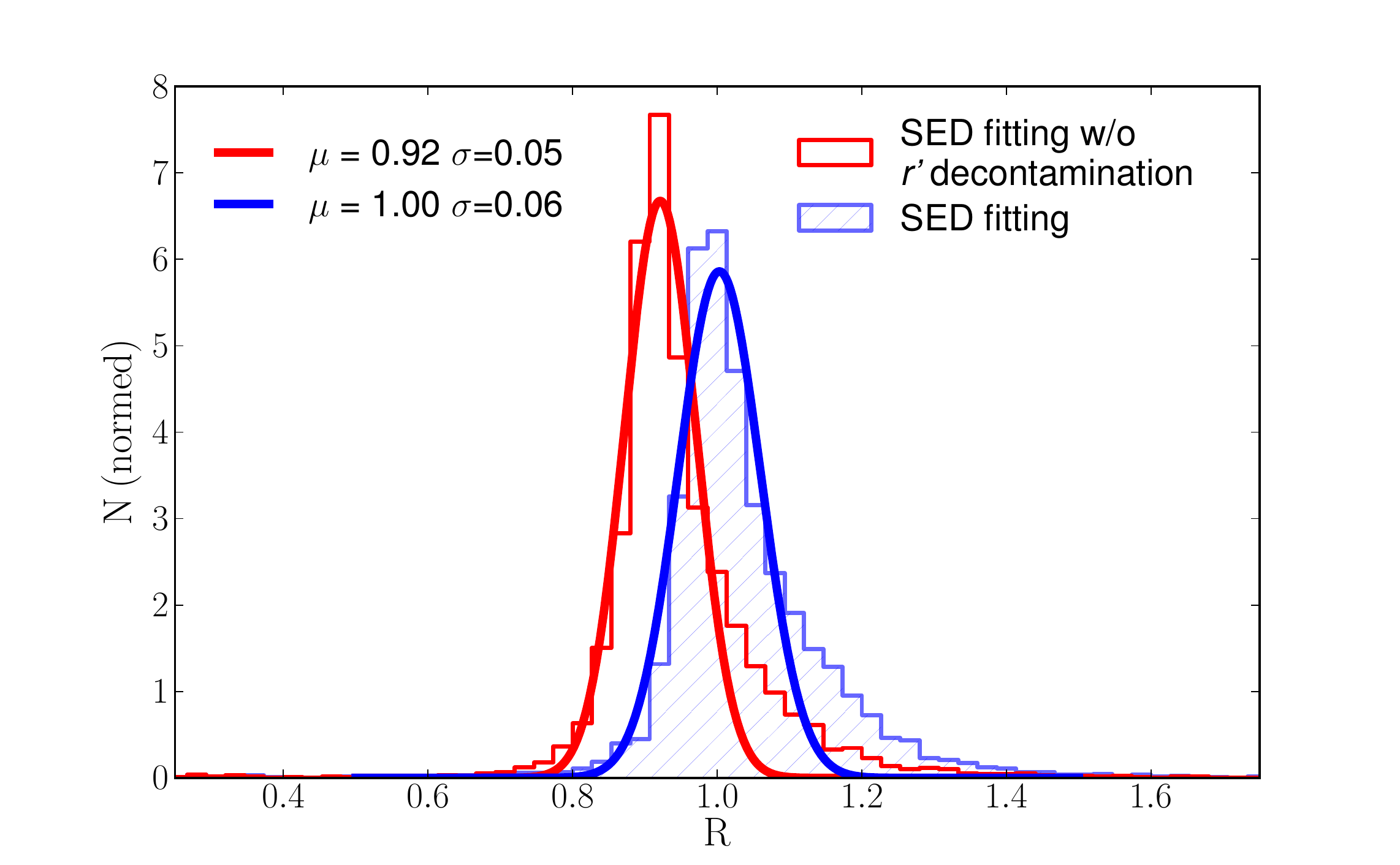}
  \caption{Normalized distribution of the ratios between the recovered and the spectroscopically measured H$\alpha$ + [\ion{N}{ii}] flux for S1 spectra, with the SED fitting routines. Empty red histogram shows the results after SED fitting method without \emph{r'} decontamination. Dashed blue histogram denotes the results after SED fitting method with \emph{r'} decontamination. Solid curves indicate Gaussian fits to the distribution of R. The best fitting values are labelled in the panel.}
  \label{gaussianasSEDS}
\end{figure}

We plot the distribution of results after applying the SED fitting methods in Fig.~\ref{gaussianasSEDS}. We compare the results without applying the \emph{r'} decontamination. The SED fitting routine with $r'$ decontamination performs better than the SED independent methodologies (i.e. the 2F and 3F methods), being unbiased ($\mu_{\mathrm{SED}}=1.00$). With this technique, we do not approximate the continuum to any function; we use the continuum inferred from BC03 templates, which also contain an estimation for the absorption of \Halpha. Another interesting result is the impact that has the H$\alpha$ + [\ion{N}{ii}] emission inside the \emph{r'} filter. Not taking this into account causes a bias in our results by $\sim8\%$. 

\subsection{Testing the methodologies: conclusions}\label{simulaciones - conclusion}
At this point we conclude that the methodologies that only use two or three filters are inconvenient for J-PLUS, given our filter configuration, as they produce biased results in average. In the case of the 2F methodology with no colour correction, this bias is $\sim22\%$, while the 3F method underestimates the \Halpha+ [\ion{N}{ii}] flux by 9\%. As \cite{Sobral2012} points out, this effect when using only two filters is, in part, caused because the central wavelength of the narrow filter does not coincide with the broad-band central wavelength. To cope with this, we need more information, and so we include the $i'$ filter. However, this does not solve totally the problem.
\newline
\indent With the SED fitting procedure we avoid this bias, though attention must be paid to the contribution of H$\alpha$ + [\ion{N}{ii}] inside the \emph{r'} filter. Not taking this effect into account biases our results by $8\%$. Finally, the SED fitting procedure with the \emph{r'} decontamination is unbiased and has a dispersion of $\sigma_{\mathrm{SED}}=0.06$. This dispersion is a combination of the errors associated with the photometry, $\delta_{\mathrm{phot}}$, and other factors that are discussed in detail in Sec.~\ref{Estimacion Errores}.\newline

\section{SED fitting routine: performance and error budget}\label{simuls_diferentes_mags}
In the previous section, we conclude that the SED fitting routine is the most reliable methodology for our purposes given the J-PLUS filter configuration. In this section, we perform some additional tests on this methodology and explore the error budget in the measurements.

\subsection{Dependence on $\mathrm{m}_{r'}$}\label{simulaciones - r_m}

\begin{figure}
  \centering
  \includegraphics[width=0.5\textwidth]{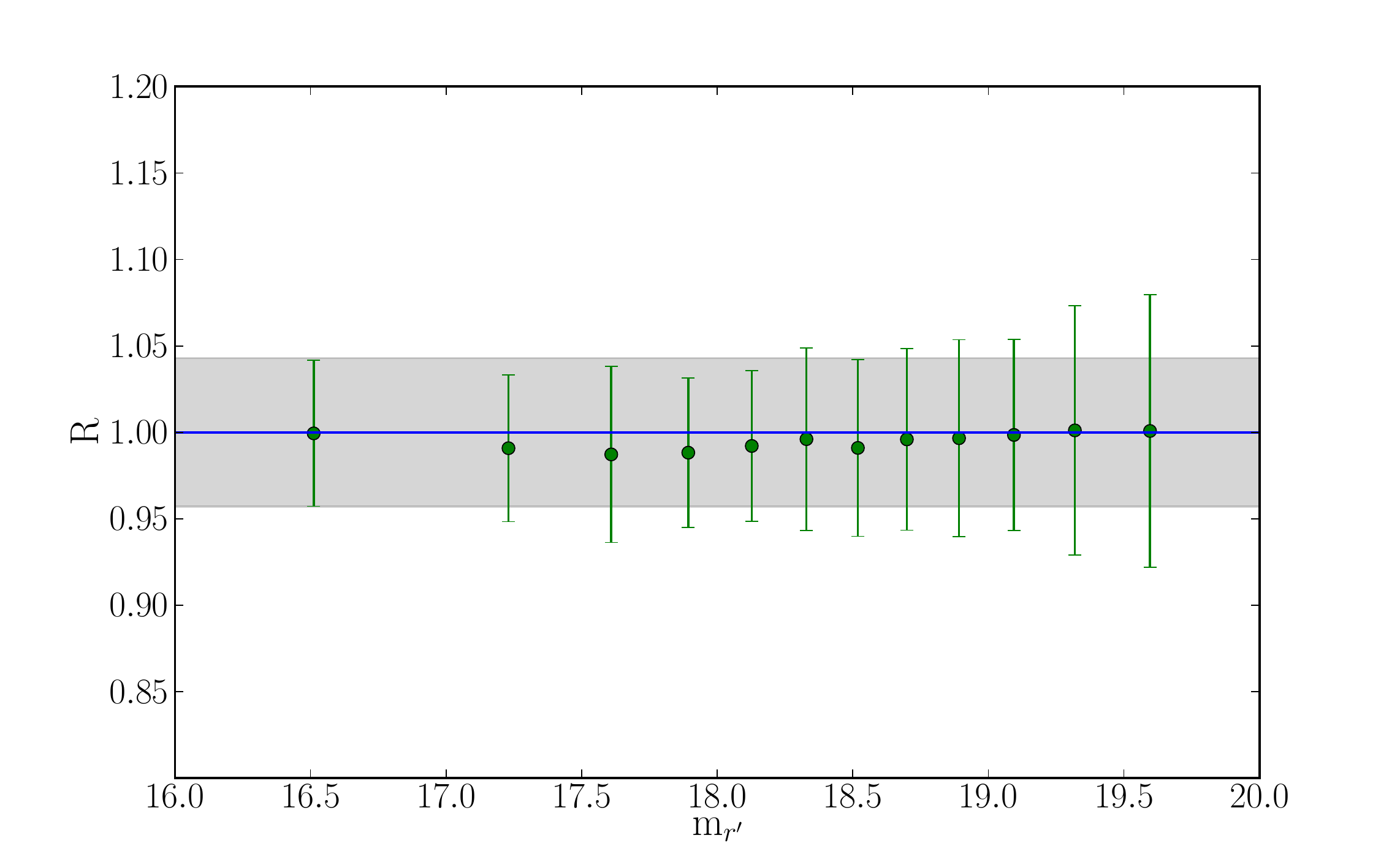}
  \caption{Green dots denote the medians of the Gaussian fitting to the galaxies inside the magnitude bin which median is $\mathrm{m}_{r'}$; Green bars indicate $\sigma$ of this Gaussian fit; The shaded area is the $4.3\%$ uncertainty defined by $\delta_{\mathrm{syst}}$.}
  \label{gaussianasSEDS_R_mags}
\end{figure}

To see if any bias appears at faint magnitudes, we study the distribution of R at several $\mathrm{m}_{r'}$ bins. To do that, we selected the galaxies in S1 within a $\mathrm{m}_{r'}$ range and fit a Gaussian to their R distribution. Bins are defined to contain the same number of galaxies, which in this case is 626.\newline 
\indent Figure~\ref{gaussianasSEDS_R_mags} shows the $\mu$ of these distributions as a function of the median $\mathrm{m}_{r'}$ magnitude of each bin. Error bars are the standard deviation $\sigma$ of each fit. The results are well recovered in all the magnitude ranges. Error bars increase from  $\sigma\sim5\%$ at $\mathrm{m}_{r'}=15$ to $\sigma\sim8\%$ at $\mathrm{m}_{r'}=19.5$. We interpret this dispersion as the combination of two effects. We see a more detailed study of these dispersions in Sect.~\ref{Simulaciones propiamente} and Sect.~\ref{Estimacion Errores}.

\subsection{Dependence on EW}\label{simulaciones - r_ew}

We repeat the same analysis, but binning our S1 as a function of the observed $\mathrm{EW}_{F660}$. This EW is not the one that we used in the data selection criteria, but that resulting from the recovered flux of H$\alpha$ + [\ion{N}{ii}],

\begin{equation}
 \mathrm{EW}_{F660}=\frac{  \langle F_{\mathrm{H}\alpha + \mathrm{[\ion{N}{ii}]}}\rangle}{F_{F660}- \langle F_{\mathrm{H}\alpha + \mathrm{[\ion{N}{ii}]}}\rangle}.
\end{equation}

\begin{figure}
  
  \centering
  \includegraphics[width=0.5\textwidth]{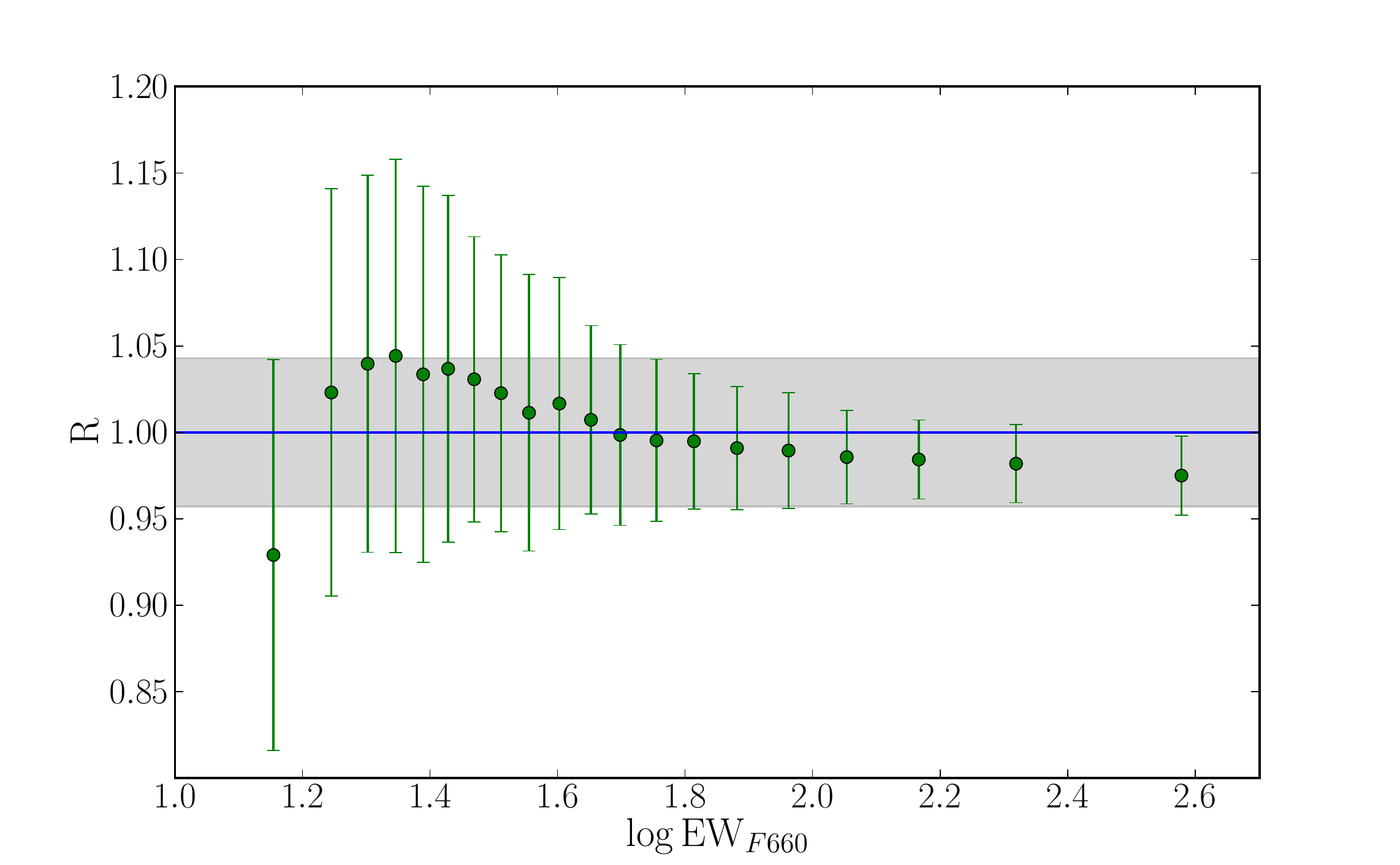}
  \caption{Green dots represent the medians of the Gaussian fitting to the galaxies inside the EW bin which median is EW; Green bars denote the $\sigma$ of the Gaussian fitting; Shaded area shows the $4.3\%$ uncertainty defined by $\delta_{\mathrm{syst}}$.}
  \label{gaussianasSEDS_R_EW}
\end{figure}

Results are shown in Fig.~\ref{gaussianasSEDS_R_EW}. Each bin contains 375 spectra. In this case, the error weighted median is still unbiased, though the results show a trend in the recovered flux that creates an excess in the region of small EW, and an underestimation in the larger EWs. This latter effect might be due to a bad determination of the continuum in regions dominated by ionized gas, where SSP models are not valid. Nevertheless, the error is constrained to under $4.3\%$.

\subsection{Simulating observations at higher $\mathrm{m}_{r'}$}\label{Simulaciones propiamente}

We have shown that observing at different $\mathrm{m}_{r'}$ does not introduce any biases. Our spectroscopic sample however cannot reach the expected J-PLUS $\mathrm{m}_{r'}$ limiting magnitude (the selection criteria naturally cut the sample before $\mathrm{m}_{r'}=20.5$, as seen in Fig.~\ref{muestra_ew}). To study the performance of the SED fitting method at magnitudes higher than $\mathrm{m}_r'=20.5$, we have to simulate observations.

\subsubsection{Sample S2}\label{SampleS2}
We selected a subsample of galaxies with $\mathrm{m}_{r'}\leq17.5$ and median ${\mathrm{S/N}}\geq 20$ from the parent sample , which were artificially scaled at any magnitude. These thresholds were chosen to have a subsample of galaxies with good quality data and a reasonable number of spectra to do statistics. In particular, we selected high S/N galaxies because we want J-PLUS photometric errors to dominate over the spectroscopic errors. We end up with a subsample of 1334 galaxies, which we call Sample S2 (Fig.~\ref{muestra_ew}).
\subsubsection{Simulation routine}\label{Simulaciones propiamente rutina}
The spectra of S2 are scaled to any $\mathrm{m}_{r'}$ of interest. After we change the magnitude of each filter with the same difference of magnitudes that we apply to match the intrinsic $\mathrm{m}_{r'}$  with the desired $\mathrm{m}_{r'}$, we compute the expected S/N for each filter, and we perturb each of the \emph{x} fluxes within its error bar from a random normal distribution with $\mu=\overline{F}_{x}$ and ${\sigma}=\nicefrac{\overline{F}_{x}}{\mathrm{S/N}_x}$. These perturbed fluxes are now considered the observations, and from these fluxes we estimate $\langle F_{\mathrm{H}\alpha + \mathrm{[\ion{N}{ii}]}}\rangle$ and its uncertainty following the process described in Sect.~\ref{Mediana y sigmaphot}.\newline

\begin{figure}
  \centering
  \includegraphics[width=0.5\textwidth]{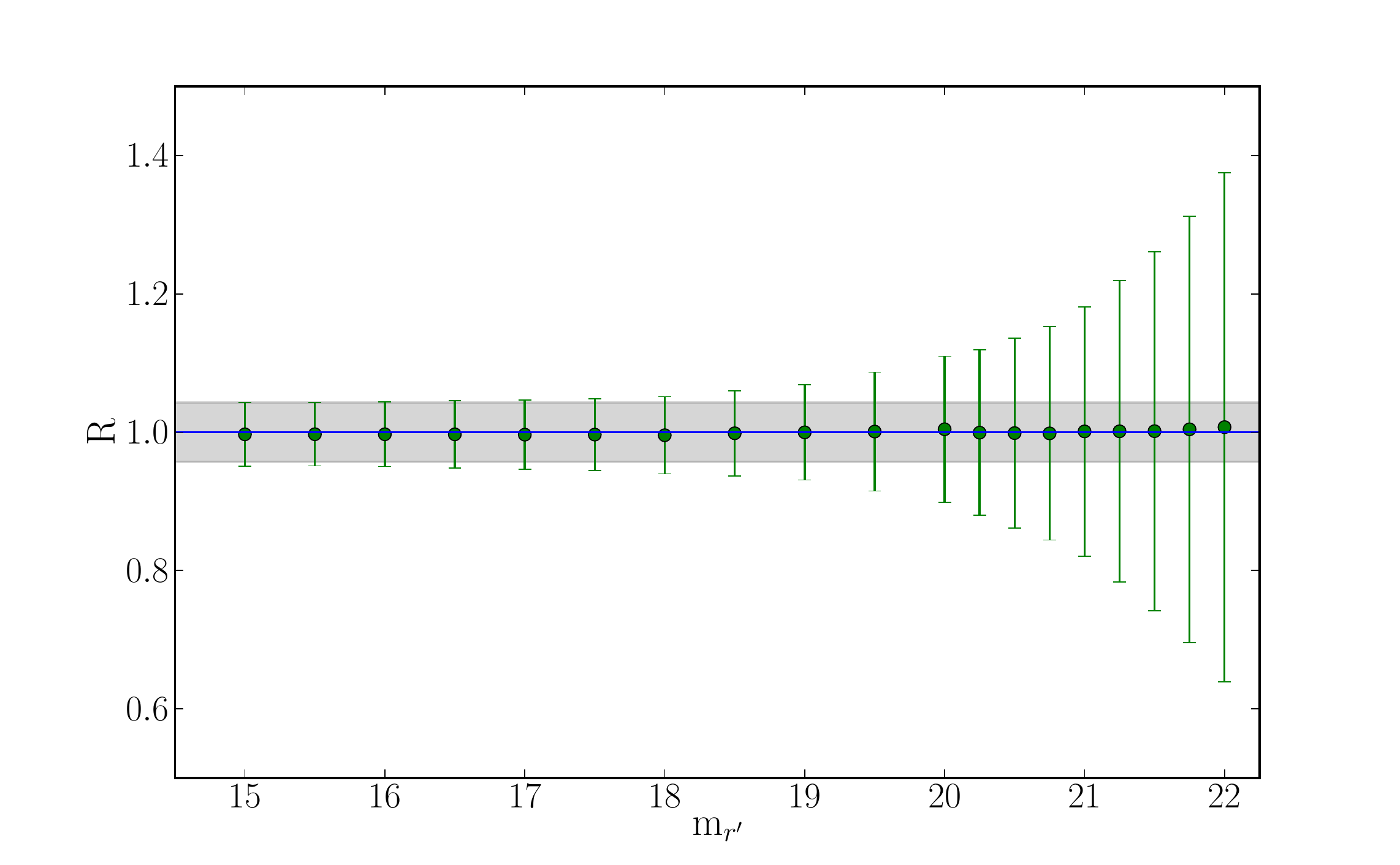}
  \caption{Results after simulating observations of S2 spectra. Green dots indicate the medians of the Gaussian fitting to the galaxies; Green bars denote the $\sigma$ of this Gaussian fitting; Shaded area illustrates $4.3\%$ uncertainty defined by $\delta_{\mathrm{syst}}$.}
  \label{gaussianasSEDS_R_mag_simuls}
\end{figure}

\indent Figure~\ref{gaussianasSEDS_R_mag_simuls} shows that we recover the fluxes without biases up to magnitude $\mathrm{m}_{r'}\sim21.8$. The errors increase at fainter magnitudes, while remains constrained at a $\sim4.3\%$ at brighter magnitudes. We discuss this in the next section.

\subsection{Estimating the errors}\label{Estimacion Errores}
Here we carry out an analysis of the errors associated with the measurement method. For convenience, we refer to the dispersion of the results, given by the standard deviation of the Gaussian fits, with letter $\sigma$; we reserve letter $\delta$ for the errors and uncertainties in the measurements. In Sect.~\ref{Mediana y sigmaphot} we explained how we estimate the photometric errors $\delta_{\mathrm{phot}}$ of  $\langle F_{\mathrm{H}\alpha + \mathrm{[\ion{N}{ii}]}}\rangle$. However, in Fig.~\ref{gaussianasSEDS_R_mag_simuls} we see that the dispersion of the results in the brightest magnitudes is almost constant. This means that we cannot explain the dispersion of the results only with $\delta_{\mathrm{phot}}$, and that we need to add a new uncertainty to our measurements, i.e.

\begin{equation}\label{sigmaphot mas sigmasist}
  \sigma_{\mathrm{SED}}=\sqrt{\delta_{\mathrm{phot}}^2+\delta_{\mathrm{syst}}^2}\:.
\end{equation}

\begin{figure}
  \centering
  \includegraphics[width=0.5\textwidth]{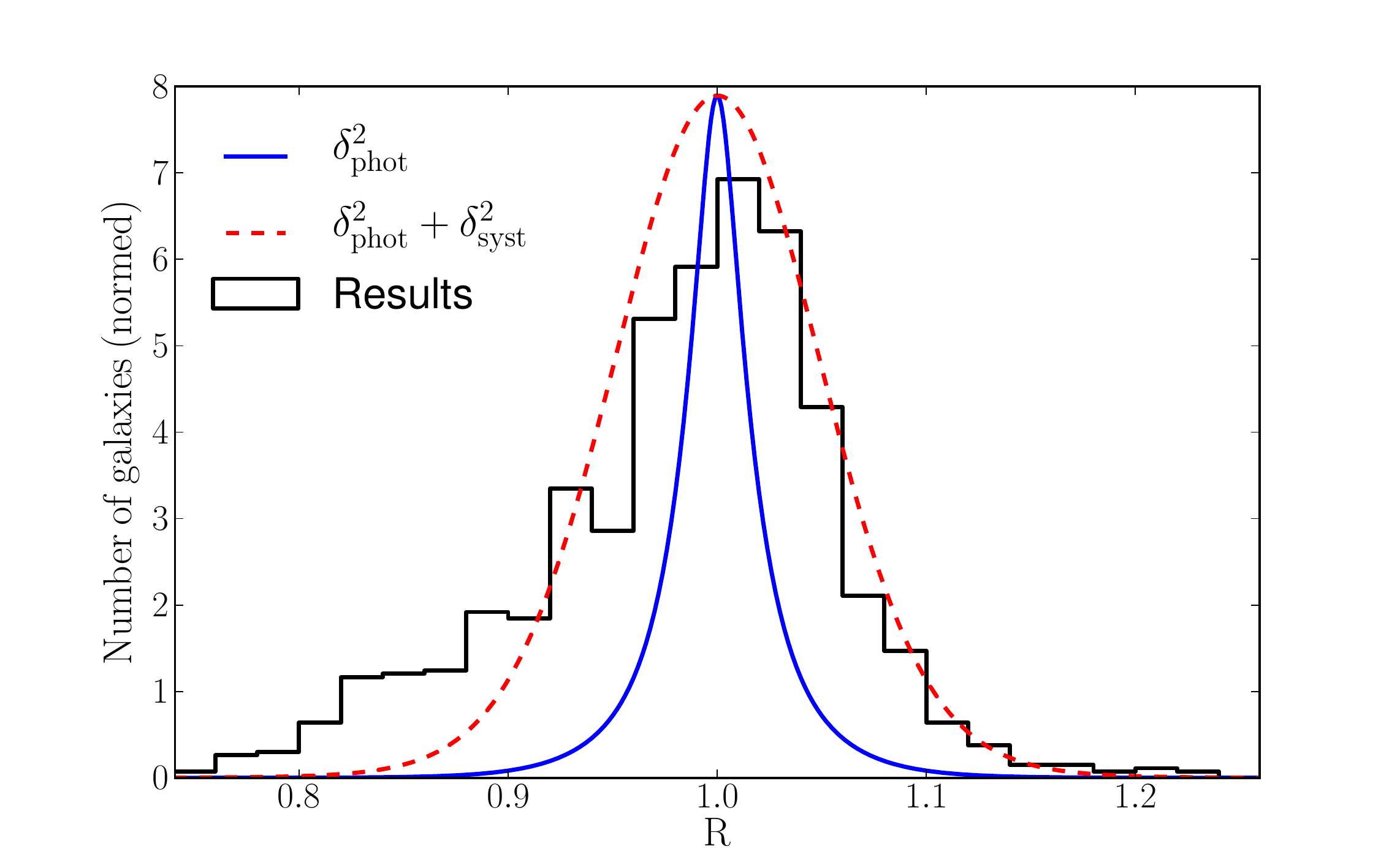}
  \includegraphics[width=0.5\textwidth]{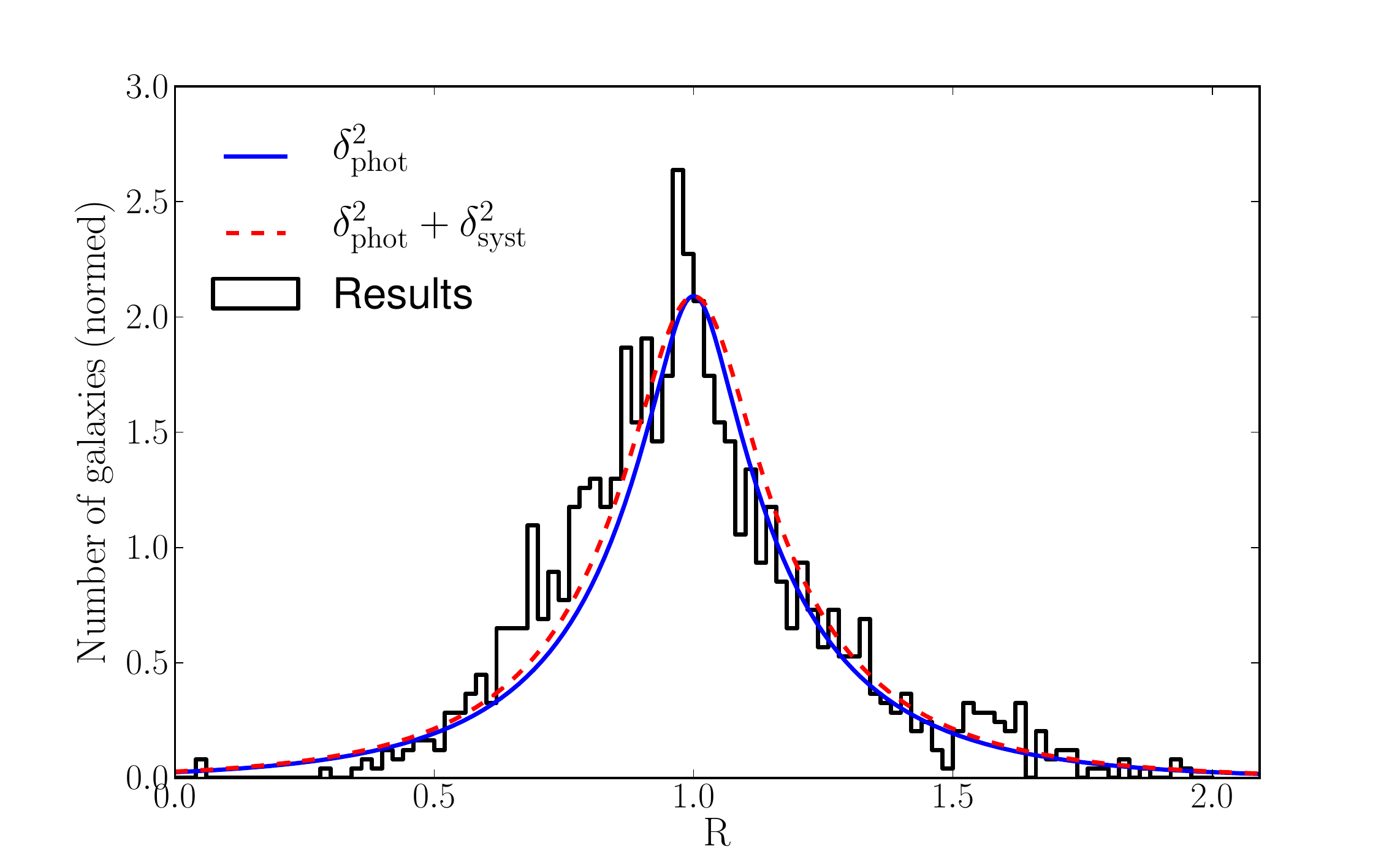}
  \caption{\textbf{Top panel}: black histogram denotes the distribution of R for the 1134 galaxies in S2, scaled to $\mathrm{m}_{r'}=17.5$. Blue curve shows the sum of 1134 normal distributions centred at $\mu=1$ and with $\sigma=\delta_{\mathrm{phot}}$. Red dashed curve is the sum of 1134 normal distributions centred at $\mu=1$ and with $\sigma=\sqrt{\delta_{\mathrm{phot}}^{2}+\delta_{\mathrm{syst}}^{2}}$. \textbf{Bottom panel}: same as above, except for spectra scaled to $\mathrm{m}_{r'}=21.25$. }
  \label{supergaussiana 1}
\end{figure}

To compute the value of $\delta_{\mathrm{syst}}$, we want to minimize $\delta_{\mathrm{phot}}$. We simulate observations with $\mathrm{S/N}>10^{8}$ for S2. In this case, the dispersion in the results is only due to $\delta_{\mathrm{syst}}$. Doing this, we find that $\delta_{\mathrm{syst}}=0.05$. The shaded area in Figs.~\ref{gaussianasSEDS_R_mags}, \ref{gaussianasSEDS_R_EW} and \ref{gaussianasSEDS_R_mag_simuls} shows this systematic uncertainty. We see that this value well constrains the error bars until these begin to increase because of photometric errors. \newline
\indent To test the validity of this result, we compare this new error with the purely photometric error. To do that, following the Monte Carlo approach described in Sect.~\ref{Mediana y sigmaphot}, we create 1134 normal distributions (one per spectrum in S2), centred in $\mu_{i}=1$ and with $\sigma_{i}=\delta_{\mathrm{phot,\,i}}$. We see in Fig.~\ref{supergaussiana 1} that, when we add all these distributions, we recover a new distribution whose dispersion is only due to photometric errors (blue solid line). However, if we repeat the same exercise with $\sigma_{i}=\sqrt{\delta_{\mathrm{phot},\:i}^2+\delta_{\mathrm{syst}}^2}$, we see that the resulting distribution has a dispersion that resembles that of our results (red dashed line).

\begin{figure}
  \centering
  \includegraphics[width=0.5\textwidth]{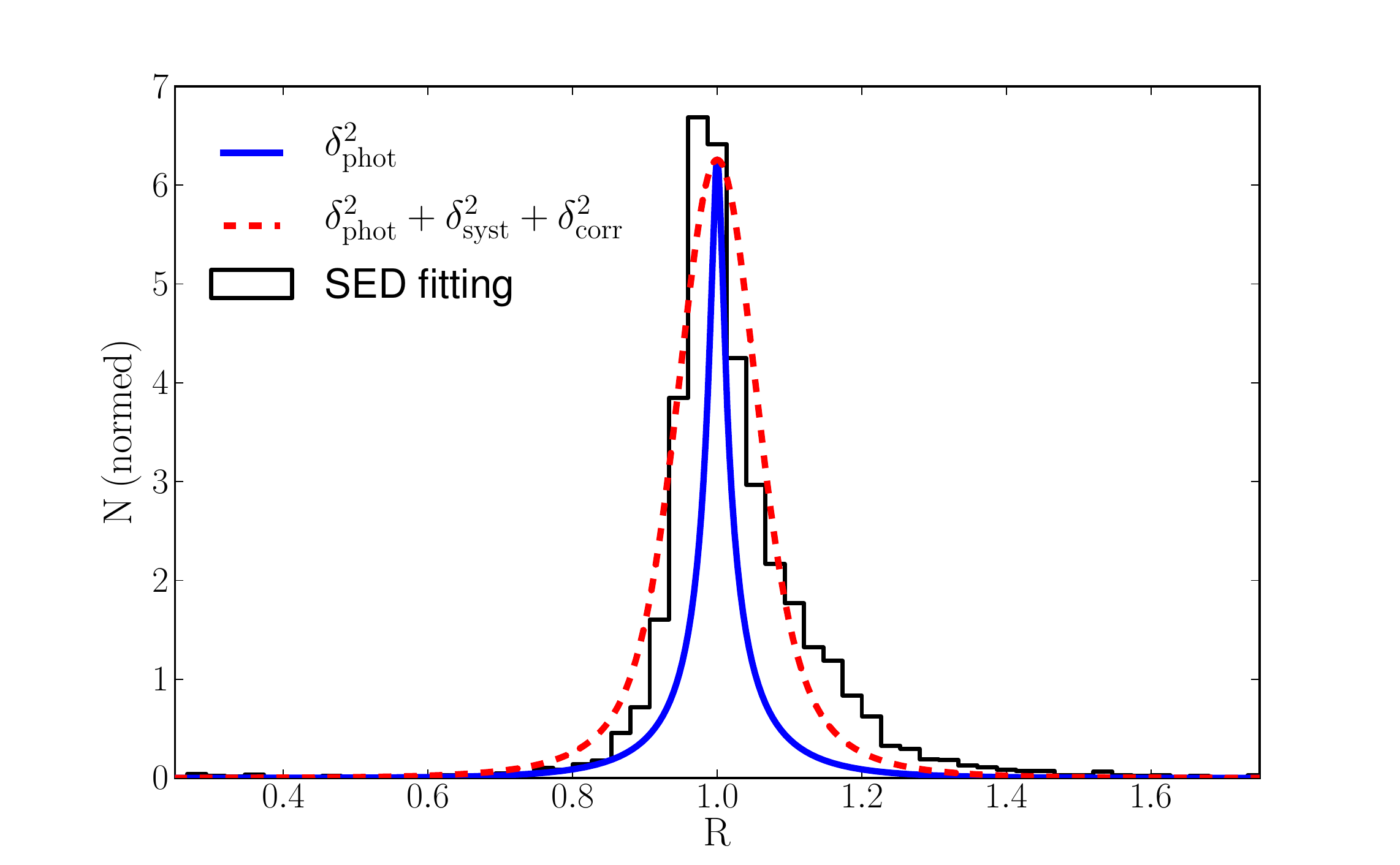}
  \caption{ Distribution of R for all the 7511 galaxies in S1. Blue curve denotes the sum of 7511 normal distributions centred at $\mu=1$ and $\sigma=\delta_{\mathrm{phot}}$. Red dashed curve indicates the sum of 7511 normal distributions centred at $\mu=1$ and with $\sigma=\sqrt{\delta_{\mathrm{phot}}^{2}+\delta_{\mathrm{syst}}^{2}}$.}
  \label{SampleS1 con errores}
\end{figure}

Finally, we apply this systematic error to the 7511 spectra of S1. We show in Fig.~\ref{SampleS1 con errores} the same distribution as in Fig.~\ref{gaussianasSEDS}, but we overplot the distribution when only taking  $\delta_{\mathrm{phot}}$ (blue solid line) into account and when we add the $4.3\%$ uncertainty. This new distribution (red solid line) traces the dispersion of our results well, meaning that our error budget is reliable.

\subsection{SED fitting routine: performance and error budget conclusions}
In this section, we have studied the sources of error that affect our measurements of H$\alpha$ + [\ion{N}{ii}] flux when using the SED fitting method. We have seen that our measurements are not biased at any magnitude of the J-PLUS detection magnitude range. To accomplish this, we studied S1 spectra at their magnitudes and simulated observations of S2 galaxies at magnitudes in which we have no data. With this test, we find that there is an uncertainty in all the magnitudes that is independent of the photometric errors. \newline
\indent We studied this uncertainty and treated it as a source of error. We see that adding a systematic uncertainty of $4.3\%$ to the \SigmaPhot of each measurement, allows us to recover a distribution that reproduces the observed dispersion in the results at any magnitude. This uncertainty can be due to a combination of several sources, such as not taking the intrinsic errors of the SDSS spectra into account, fitting SSPs to regions that may not be well represented by SSPs, or differences in the measurement procedures between our results and the Portsmouth Group.

\section{Dust correction and [\ion{N}{ii}] removal}\label{Polvo y nitrógenos}
\subsection{Dust correction}
Our aim is to recover the \Halpha emission from galaxies; however, galaxies contain dust that is mixed with the stellar populations and the gas. Dust is present in molecular clouds before they collapse to form stars, and it is mixed with the hot gas after the first stars of a star-forming region are born. The presence of dust has two important observable consequences: it attenuates the total amount of light that we receive, and tends to redden the true colour of light-emitting region. Both effects have an impact on the absolute and relative fluxes and magnitudes that we measure. Attenuation causes all the magnitudes in the optical and UV to increase, while reddening causes that this increase is higher as we move to the blue parts of the spectrum. \newline
\indent The proportion between the attenuation in the Johnson V band ($A_{V}$) and the difference between the observed and the intrinsic (i.e. dust-free) magnitudes of the Johnson B and V bands (\emph{i.e.} the colour excess $E(B-V)$) has been called the extinction law \citep[see][]{Cardelli1989}, although definitions in other bands exist \citep{Fitzpatrick1999}, i.e. 

\begin{equation}
 R_{V}=\frac{A_{V}}{E(B-V)}.
\end{equation}

From \cite{Calzetti2000}, we know that the relation between the intrinsic flux and the observed one is

\begin{equation}
 F_{i}(\lambda)=F_{o}(\lambda)10^{0.4A_{\lambda}}=F_{o}(\lambda)10^{0.4E(B-V)k'(\lambda)}
\end{equation}

\noindent where $k'(\lambda)$ is a polynomial that depends on $\lambda$ and $R_{V}$.

To compare the recovered flux with the one provided by the Portsmouth Group, we added dust to their measurements with the values of $E(B-V)$ that they provided (see Eq.~\ref{polvoanadido} ). However, with the filter configuration of J-PLUS it is more difficult to estimate the dust contribution to our fluxes.
\newline
\indent From now on we will be using the SED fitting procedure to extract the flux of H$\alpha$ + [\ion{N}{ii}], as the other methodologies presented biases. We stress that the corrections presented in this section can be modified in the future because they are independent of the method that we used to isolate the emission lines. \newline

\begin{figure}
  
  \centering
  \includegraphics[width=0.5\textwidth]{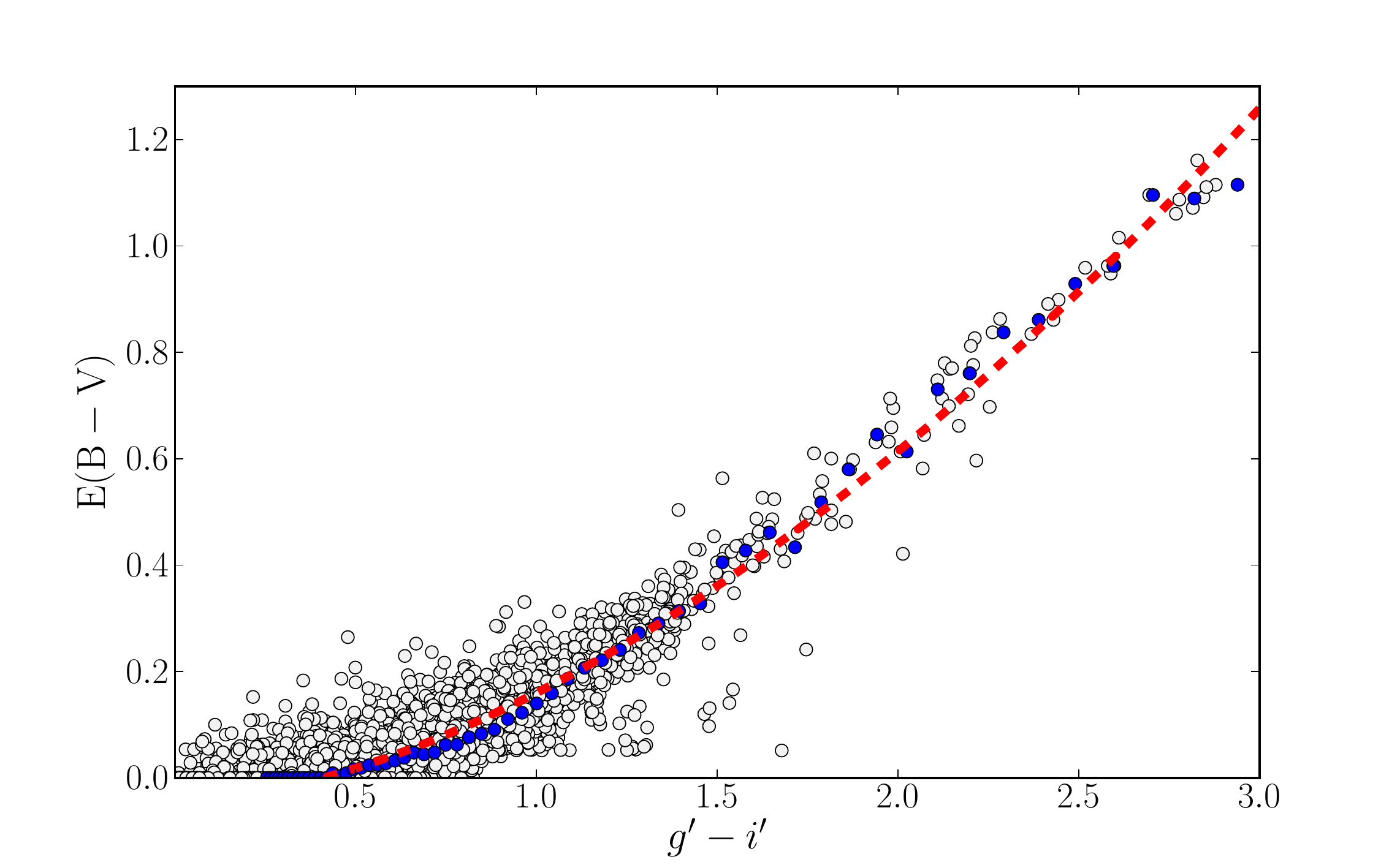}
  \caption{Distribution of spectroscopically derived $E(B-V)$ and observed $g'-i'$. Blue dots denote the medians of $E(B-V)$ values enclosed in the logarithmically spaced bins. Red dashed line shows the fit to the blue dots.}
  \label{ebmv_gi}
\end{figure}

A common assumption that is applied to photometric data is an attenuation of $A_{\mathrm{H}\alpha}=1\:\mathrm{mag}$ for the H$\alpha$ emission \citep[see][]{Kennicutt1992,Geach2008,Villar2008,Sobral2012}. The dust-corrected values assuming this attenuation are $F_{\mathrm{H}\alpha+\mathrm{[\ion{N}{ii}]}}=F_{F660}\times10^{0.4A_{\mathrm{H}\alpha}}=2.5F_{F660}$. However, we applied this correction and found that it tends to overestimate the results, being the median of R $\mu=2.3$, far from the expected $\mu=1$. \newline
\indent Other studies, such as \cite{Sobral2014}, use more sophisticated techniques to correct dust extinction. In their work, the correction by \cite{Garn2010} for star-forming galaxies is applied. This correction relates dust extinction with the stellar mass of galaxies. However, we do not aim to derive stellar masses for our test sample, as at the redshift range of our interest the SDSS spectra generally do not cover the entire galaxies, but a small portion of them is covered by the SDSS fibre. \newline
\indent With S1 we find that there is a trend between the observed $g'-i'$ colour and the spectroscopically measured $E(B-V)$. This is represented in Fig.~\ref{ebmv_gi}. To fit a power-law function to these points, we bin this sample in $g'-i'$ logarithmically spaced bins and compute the median of each bin. We see in Fig.~\ref{ebmv_gi} that, for $g'~-~i'\lesssim0.5$, medians of $E(B-V)$ are 0. This is because there is a subsample of data, which has $E(B-V)=0$ and no associated error. We take them into account as zero for the fit and we obtain

\begin{equation}\label{EcPolvo}
  E(B-V) =  0.206\left(g'-i'\right)^{\:1.68} -0.0457.
\end{equation}

\indent We avoid negative colour excess values by making $E(B~-~V)=0$ if the inferred value is negative.  The derived relation should only be used in certain circumstances. The $g'-i'$ colour that we are using here is obtained after convolving the SDSS spectra with the J~-PLUS photometric system. This means that this colour is a local property of the region inside the SDSS fibre. With J-PLUS we will have spatially resolved galaxies, where we will be able to differentiate and isolate star-forming regions. These are the regions where this relation is reliable. For galaxies that are not spatially resolved, the $g'-i'$ colour is an overall colour resulting of the underlying stellar populations and the gas, if any. In these cases, the validity of this correction is not ensured, and other SED-fitting codes which study galaxy properties and stellar populations in more detail, may be used to explore dust extinction \citep[see, for instance, \emph{MUFFIT},][]{MUFFIT}.

\subsection{[\ion{N}{ii}] correction}
The $F660$ filter contains the flux of H$\alpha$ and [\ion{N}{ii}] doublet, and it is not possible to deblend these three lines to isolate the H$\alpha$ emission flux. To cope with this problem, empirical relations must be applied. With data from S1, we find that there is a relation between the flux of \Halpha + [\ion{N}{ii}] and the \Halpha flux alone. Figure~\ref{Nitrogeno} shows this relation. We find that there is a slight bimodality in the distribution of fluxes, which is blurred in the low-emission regime. This bi-modality can be disentangled with the help of the colour $g'-i'$, and we fit the following equation to each branch:

\begin{equation}\label{EcNitrogenos}
 \log(\mathrm{H}\alpha) =  \begin{cases}
   0.989\log(F_{\mathrm{H}\alpha+\mathrm{[\ion{N}{ii}]},\:\mathrm{D.C.}})-0.193,  & \text{if } g'-i'  \leq 0.5, \\
   0.954\log(F_{\mathrm{H}\alpha+\mathrm{[\ion{N}{ii}]},\:\mathrm{D.C.}})-0.753,  & \text{if } g'-i' > 0.5,
   
  \end{cases}
\end{equation}

\noindent where $F_{\mathrm{H}\alpha+\mathrm{[\ion{N}{ii}]},\:\mathrm{D.C.}}$ refers to the $F660$ flux after dust correction. 

\begin{figure}
  
  \centering
  \includegraphics[width=0.5\textwidth]{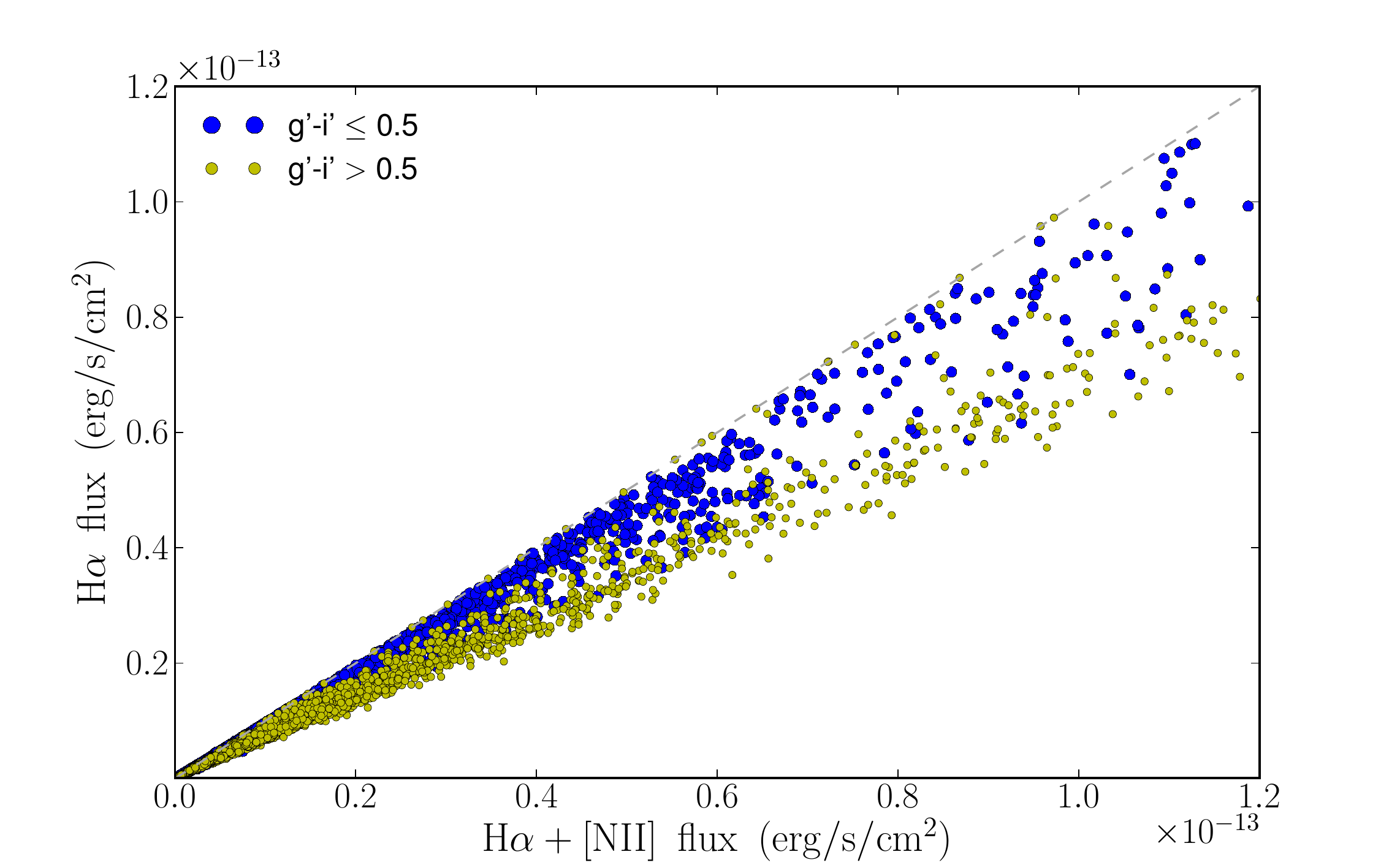}
  \caption{Relation between the spectroscopically measured and dust corrected $\mathrm{H}\alpha$ flux and the total H$\alpha$ + [\ion{N}{ii}] flux. We see that the distribution is bimodal. The two trends can be differentiated if we split the sample by its observed $g'-i'$ colour.}
  \label{Nitrogeno}
\end{figure}

\subsection{\Halpha only measurements}

We apply both corrections to our measurements and compare the recovered results with the spectroscopic values of \Halpha+[\ion{N}{ii}] without dust and \Halpha only. As we can see in Fig.~\ref{Nitrogeno}, both corrections help us to recover the $\mathrm{H}\alpha$ flux without adding any extra bias to the whole set of measurements. To these results, we fit a Gaussian distribution.

\begin{figure}
  \centering
  \includegraphics[width=0.5\textwidth]{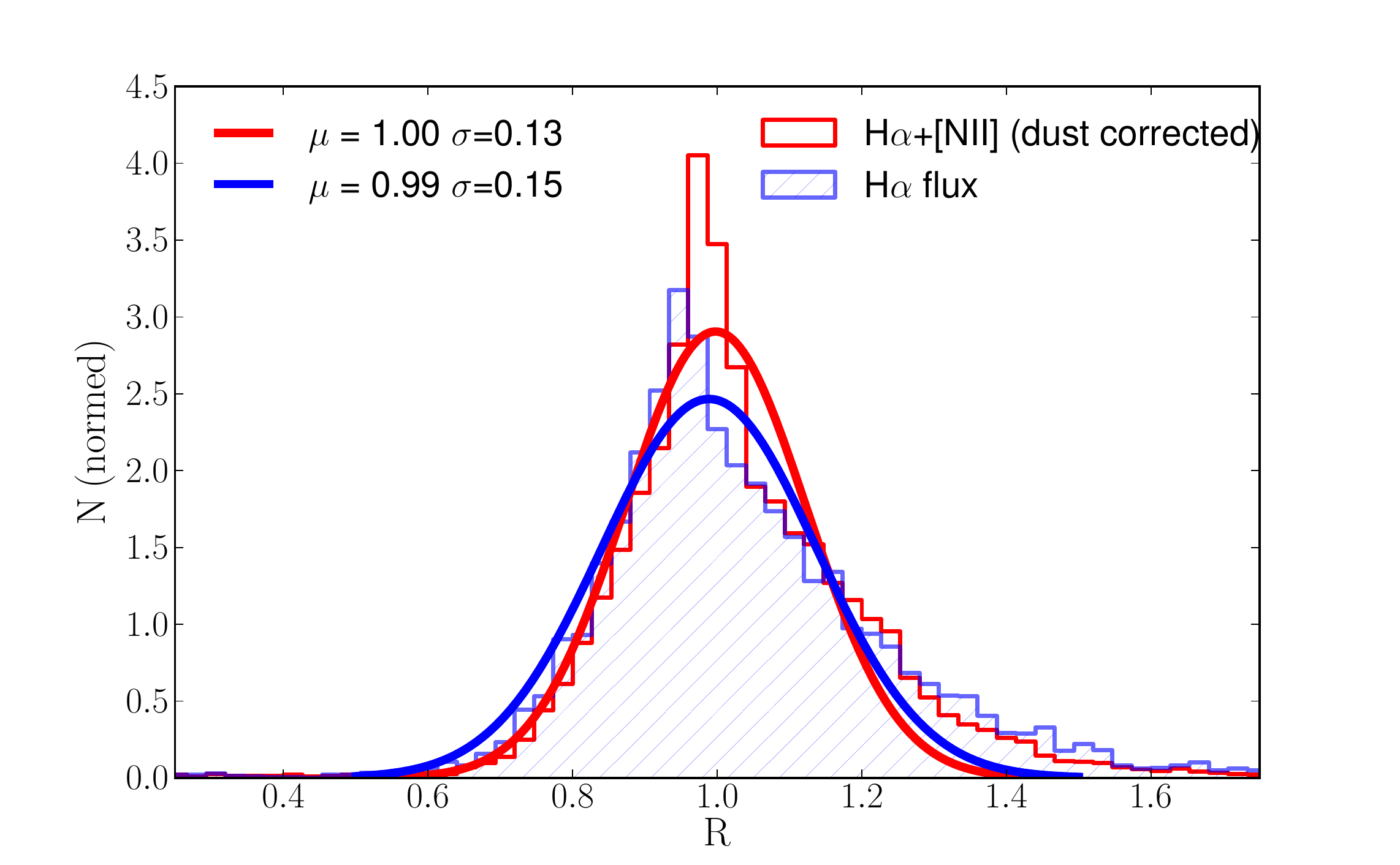}
  \caption{The empty red histogram is distribution of the recovered \Halpha + [\ion{N}{ii}] after correcting for dust; dashed blue histogram denotes the distribution of the recovered \Halpha flux after correcting for dust and [\ion{N}{ii}]; Solid lines indicate Gaussian fits to the data. Best-fitting values are labelled in the panel.}
  \label{Nitrogeno}
\end{figure}

It is important to stress that [\ion{N}{ii}] correction is empirical, and can only be applied after correcting for dust the observed $F660$ flux. This relation should hold regardless of the dust correction that is applied, as it has been calibrated with dust-free data.

\subsection{Error budget}\label{Error budget correcciones}
After correcting the flux of H$\alpha$ + [\ion{N}{ii}] from dust, the resulting distribution has a larger dispersion, although it does not become biased. This increase in the dispersion must be taken as another source of uncertainty, namely $\delta_{\mathrm{corr}}$, i.e.

\begin{equation}\label{sigma con sisttematico y polvo y nitrogeno}
 \delta_{\mathrm{H}\alpha}=\sqrt{\delta_{\mathrm{phot}}^2+\delta_{\mathrm{syst}}^2+\delta_{\mathrm{corr}}^2}\:.
\end{equation}

To derive the value of $\delta_{\mathrm{corr}}$, we compare the dispersion in distribution of raw \Halpha + [\ion{N}{ii}] ($\sigma_{\mathrm{SED}}=0.06$, Fig.~\ref{gaussianasSEDS}), only with the dispersion of the distribution of \Halpha ($\sigma_{\mathrm{H}\alpha}=0.15$, Fig.~\ref{Nitrogeno}). We find that $\delta_{\mathrm{corr}}=0.14$.

Combining both uncertainties $\delta_{\mathrm{syst}}$ (Sect.~\ref{Estimacion Errores}) and $\delta_{\mathrm{corr}}$, we obtain that the final error in \Halpha is

\begin{equation}\label{sigma con sisttematico y polvo y nitrogeno}
 \delta_{\mathrm{H}\alpha}=\sqrt{\delta_{\mathrm{phot}}^2+0.15^2}\:.
\end{equation}

\begin{figure}
  \centering
  \includegraphics[width=0.5\textwidth]{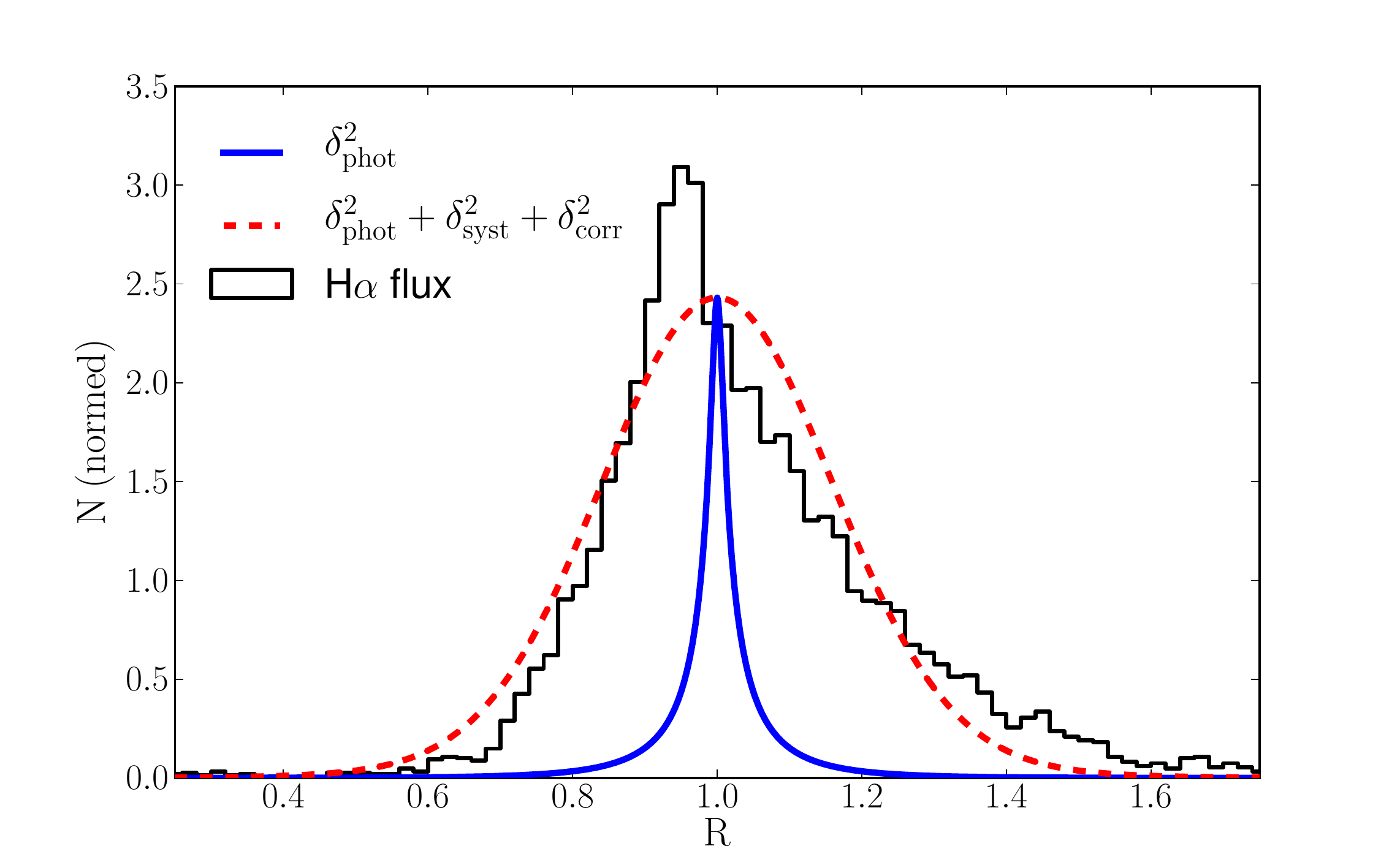}
  \caption{ The Histogram is the distribution of the recovered \Halpha flux after correcting for dust and [\ion{N}{ii}]; The solid curve shows the distribution only when taking $\delta_{\mathrm{phot}}$ into account; the dashed curve shows the distribution when adding $\delta_{\mathrm{syst}}$ and $\delta_{\mathrm{corr}}$ to $\delta_{\mathrm{phot}}$.}
  \label{Nitrogeno con errores}
\end{figure}

We now add this error to each measurement of S1 spectra and repeat the same analysis as in Sect.~\ref{Estimacion Errores}. The resulting distribution, with the fitting properties and the errors, is shown in Fig.~\ref{Nitrogeno con errores}. We see that a $15\%$ uncertainty creates a distribution that resembles the dispersion of R after correcting for dust and [\ion{N}{ii}], as desired.

\subsection{Dust correction and [\ion{N}{ii}] removal: conclusions}
In this section, we have studied how to correct for dust extinction and for the contribution of the [\ion{N}{ii}] doublet to the total observed flux inside $F660$. We used two empirical relations derived from SDSS data. Taking the properties of galaxies used to derive these expressions into account, they should be representative of the properties of star-forming regions of galaxies in the local Universe.\newline 
\indent When applied, both corrections retrieve unbiased results. This means that we can decouple them from our first goal: obtaining reliable measurements of H$\alpha$ + [\ion{N}{ii}]. In this sense, both corrections can be modified in the future if better corrections are found.\newline
\indent When we add both corrections to the raw measurement of H$\alpha$ + [\ion{N}{ii}], the distribution of results is unbiased, but the dispersion increases. We obtain that the uncertainty introduced when we apply our corrections is $\delta_{\mathrm{corr}}\sim14\%$.

\section{Summary and conclusions}\label{summary}

We have presented the capabilities of the J-PLUS survey to infer H$\alpha$ emission from photometric data. We first presented different methodologies and equations that can be applied to extract emission line fluxes from narrow-band and broad-band photometry.  After that, we tested each methodology simulating observations of SDSS spectra as seen by J-PLUS. We find that:
\begin{enumerate}
 \item Using a broad- and a narrow-band filter without taking care of the colour of the galaxy retrieves severely biased results, tending to underestimate the H$\alpha$ + [\ion{N}{ii}] flux of star-forming galaxies by  $\sim20\%$. The asymmetry of the distribution makes it  difficult to treat it statistically and to have the errors under control. To use this methodology, colour corrections should be applied, as shown by \cite{Sobral2009}.
 \item Using a combination of two broad-band filters and a narrow-band gives better results. However the method is still biased in average by a 9\%.
 \item Fitting the whole SED to a collection of SSP models after subtracting the \Halpha + [\ion{N}{ii}] flux from \emph{r'}, retrieves unbiased \Halpha + [\ion{N}{ii}] fluxes. We stress the importance of taking into account the contribution of the emission lines inside the \emph{r'} filter. Not correcting the flux of this filter introduces a bias of $8\%$.
\end{enumerate}

We conclude that the SED fitting method is the best one given the J-PLUS capabilities. We encourage other photometric surveys targeting emission lines with narrow-band filters to explore the SED fitting methodology, instead of restricting them to the use of two or three filters. \newline

To correct the observed \Halpha + [\ion{N}{ii}] emission flux from dust and remove the [\ion{N}{ii}] contribution, we derived empirical corrections from the SDSS data. After that, the recovered \Halpha flux is still unbiased, but suffers a larger dispersion.

\indent Finally, we demonstrate that the error of our measurements of \Halpha flux, $\delta_{\mathrm{H}\alpha}$, has three contributions. The first is given by the photometric errors of our data, $\delta_{\mathrm{phot}}$. The second contribution has several sources, which include not considering the intrinsic errors of SDSS spectra, the use of SSPs to fit the stellar continuum of regions that may not resemble SSPs, and differences between our measurement procedure and the Portsmouth group's method. This results in a systematic uncertainty, $\delta_{\mathrm{syst}}$. The last source of uncertainty is related with the corrections of dust and [\ion{N}{ii}], namely $\delta_{\mathrm{corr}}$. In the end, we express the error of our $\mathrm{H}\alpha$  measurements as

\begin{equation}
 \delta_{\mathrm{H}\alpha}=\sqrt{\delta_{\mathrm{phot}}^2 +\delta_{\mathrm{syst}}^2+\delta_{\mathrm{corr}}^2 },
\end{equation}

\indent where $\delta_{\mathrm{syst}}=0.05$ (Sect.~\ref{Estimacion Errores}) and $\delta_{\mathrm{corr}}=0.14$ (Sect.~\ref{Error budget correcciones}). This means that our \Halpha measurements have a  $15\%$ of uncertainty, regardless of the quality of the data. 

We stress here that this $15\%$ is an upper limit to the uncertainty related to the methodology and the corrections. Because each stage of the process (emission detection, dust correction, and [\ion{N}{ii}] correction) is unbiased, we can decouple them. This would allow us to improve these corrections and reduce the systematic error. For instance, J-PLUS has a narrow filter in the [\ion{O}{ii}] wavelength range. This forbidden emission is also a tracer of the SFR, and could help us constrain the dust contribution better than the colour that we used.

\begin{acknowledgements} 
This work has mainly been funding by the FITE (Fondos de Inversiones de Teruel) and the project AYA2012-30789.
G.V.R. is funded by Spanish Ayudas para contratos predoctorales para la formaci\'on de doctores grant BES-2013-065051

This research made use of \texttt{Matplotlib}, a 2D graphics package used for \texttt{Python} for publication-quality image generation across user interfaces and operating systems \citep{pylab}. G.V. wants to thank all the science staff at CEFCA for the fruitful teachings in the Tuesdays astro-coffees, and Dr. C. Hern\'andez-Monteagudo for the help during the ascension to Pico do Papagayo (Ilha Grande, Brasil).
Funding for SDSS-III has been provided by the Alfred P. Sloan Foundation, the Participating Institutions, the National Science Foundation, and the U.S. Department of Energy Office of Science. The SDSS-III web site is http://www.sdss3.org/.

SDSS-III is managed by the Astrophysical Research Consortium for the Participating Institutions of the SDSS-III Collaboration including the University of Arizona, the Brazilian Participation Group, Brookhaven National Laboratory, Carnegie Mellon University, University of Florida, the French Participation Group, the German Participation Group, Harvard University, the Instituto de Astrofisica de Canarias, the Michigan State/Notre Dame/JINA Participation Group, Johns Hopkins University, Lawrence Berkeley National Laboratory, Max Planck Institute for Astrophysics, Max Planck Institute for Extraterrestrial Physics, New Mexico State University, New York University, Ohio State University, Pennsylvania State University, University of Portsmouth, Princeton University, the Spanish Participation Group, University of Tokyo, University of Utah, Vanderbilt University, University of Virginia, University of Washington, and Yale University. 
\end{acknowledgements}

\bibliography{bibliografiaTFM_2}

\begin{thebibliography}{40}
\expandafter\ifx\csname natexlab\endcsname\relax\def\natexlab#1{#1}\fi

\bibitem[{{An} {et~al.}(2014){An}, {Zheng}, {Wang}, {Huang}, {Kong}, {Wang},
  {Fang}, {Zhu}, {Gu}, {Wu}, {Hao}, \& {Xia}}]{An2014}
{An}, F.~X., {Zheng}, X.~Z., {Wang}, W.-H., {et~al.} 2014, \apj, 784, 152

\bibitem[{{Baldwin} {et~al.}(1981){Baldwin}, {Phillips}, \&
  {Terlevich}}]{Baldwin1981}
{Baldwin}, J.~A., {Phillips}, M.~M., \& {Terlevich}, R. 1981, \pasp, 93, 5

\bibitem[{{Benitez} {et~al.}(2014){Benitez}, {Dupke}, {Moles}, {Sodre},
  {Cenarro}, {Marin-Franch}, {Taylor}, {Cristobal}, {Fernandez-Soto}, {Mendes
  de Oliveira}, {Cepa-Nogue}, {Abramo}, {Alcaniz}, {Overzier},
  {Hernandez-Monteagudo}, {Alfaro}, {Kanaan}, {Carvano}, {Reis}, {Martinez
  Gonzalez}, {Ascaso}, {Ballesteros}, {Xavier}, {Varela}, {Ederoclite},
  {Vazquez Ramio}, {Broadhurst}, {Cypriano}, {Angulo}, {Diego}, {Zandivarez},
  {Diaz}, {Melchior}, {Umetsu}, {Spinelli}, {Zitrin}, {Coe}, {Yepes}, {Vielva},
  {Sahni}, {Marcos-Caballero}, {Shu Kitaura}, {Maroto}, {Masip}, {Tsujikawa},
  {Carneiro}, {Gonzalez Nuevo}, {Carvalho}, {Reboucas}, {Carvalho}, {Abdalla},
  {Bernui}, {Pigozzo}, {Ferreira}, {Chandrachani Devi}, {Bengaly}, {Campista},
  {Amorim}, {Asari}, {Bongiovanni}, {Bonoli}, {Bruzual}, {Cardiel}, {Cava},
  {Cid Fernandes}, {Coelho}, {Cortesi}, {Delgado}, {Diaz Garcia}, {Espinosa},
  {Galliano}, {Gonzalez-Serrano}, {Falcon-Barroso}, {Fritz}, {Fernandes},
  {Gorgas}, {Hoyos}, {Jimenez-Teja}, {Lopez-Aguerri}, {Lopez-San Juan},
  {Mateus}, {Molino}, {Novais}, {OMill}, {Oteo}, {Perez-Gonzalez}, {Poggianti},
  {Proctor}, {Ricciardelli}, {Sanchez-Blazquez}, {Storchi-Bergmann}, {Telles},
  {Schoennell}, {Trujillo}, {Vazdekis}, {Viironen}, {Daflon},
  {Aparicio-Villegas}, {Rocha}, {Ribeiro}, {Borges}, {Martins}, {Marcolino},
  {Martinez-Delgado}, {Perez-Torres}, {Siffert}, {Calvao}, {Sako}, {Kessler},
  {Alvarez-Candal}, {De Pra}, {Roig}, {Lazzaro}, {Gorosabel}, {Lopes de
  Oliveira}, {Lima-Neto}, {Irwin}, {Liu}, {Alvarez}, {Balmes}, {Chueca},
  {Costa-Duarte}, {da Costa}, {Dantas}, {Diaz}, {Fabregat}, {Ferrari},
  {Gavela}, {Gracia}, {Gruel}, {Gutierrez}, {Guzman}, {Hernandez-Fernandez},
  {Herranz}, {Hurtado-Gil}, {Jablonsky}, {Laporte}, {Le Tiran}, {Licandro},
  {Lima}, {Martin}, {Martinez}, {Montero}, {Penteado}, {Pereira}, {Peris},
  {Quilis}, {Sanchez-Portal}, {Soja}, {Solano}, {Torra}, \&
  {Valdivielso}}]{BenitezJPAS}
{Benitez}, N., {Dupke}, R., {Moles}, M., {et~al.} 2014, ArXiv e-prints

\bibitem[{{Bruzual} \& {Charlot}(2003)}]{BC03}
{Bruzual}, G. \& {Charlot}, S. 2003, \mnras, 344, 1000

\bibitem[{{Calzetti}(2013)}]{Calzetti2013}
{Calzetti}, D. 2013, {Star Formation Rate Indicators}, ed.
  J.~{Falc{\'o}n-Barroso} \& J.~H. {Knapen}, 419

\bibitem[{{Calzetti} {et~al.}(2000){Calzetti}, {Armus}, {Bohlin}, {Kinney},
  {Koornneef}, \& {Storchi-Bergmann}}]{Calzetti2000}
{Calzetti}, D., {Armus}, L., {Bohlin}, R.~C., {et~al.} 2000, \apj, 533, 682

\bibitem[{{Calzetti} {et~al.}(2007){Calzetti}, {Kennicutt}, {Engelbracht},
  {Leitherer}, {Draine}, {Kewley}, {Moustakas}, {Sosey}, {Dale}, {Gordon},
  {Helou}, {Hollenbach}, {Armus}, {Bendo}, {Bot}, {Buckalew}, {Jarrett}, {Li},
  {Meyer}, {Murphy}, {Prescott}, {Regan}, {Rieke}, {Roussel}, {Sheth}, {Smith},
  {Thornley}, \& {Walter}}]{Calzetti2007}
{Calzetti}, D., {Kennicutt}, R.~C., {Engelbracht}, C.~W., {et~al.} 2007, \apj,
  666, 870

\bibitem[{{Cardelli} {et~al.}(1989){Cardelli}, {Clayton}, \&
  {Mathis}}]{Cardelli1989}
{Cardelli}, J.~A., {Clayton}, G.~C., \& {Mathis}, J.~S. 1989, \apj, 345, 245

\bibitem[{{Cenarro} {et~al.}(2014){Cenarro}, {Moles}, {Mar{\'{\i}}n-Franch},
  {Crist{\'o}bal-Hornillos}, {Yanes D{\'{\i}}az}, {Ederoclite}, {Varela},
  {V{\'a}zquez-Rami{\'o}}, {Valdivielso}, {Ben{\'{\i}}tez}, {Cepa}, {Dupke},
  {Fern{\'a}ndez-Soto}, {Mendes de Oliveira}, {Sodr{\'e}}, {Taylor},
  {Rueda-Teruel}, {Rueda-Teruel}, {Luis-Simoes}, {Chueca}, {Ant{\'o}n},
  {Bello}, {D{\'{\i}}az-Mart{\'{\i}}n}, {Guill{\'e}n-Civera},
  {Hern{\'a}ndez-Fuertes}, {Iglesias-Marzoa}, {Jim{\'e}nez-Mej{\'{\i}}as},
  {Lasso-Cabrera}, {L{\'o}pez-Alegre}, {L{\'o}pez-Sainz},
  {Rodr{\'{\i}}guez-Hern{\'a}ndez}, {Su{\'a}rez}, {Lamadrid}, {Ma{\'{\i}}cas},
  {Abril-Iba{\~n}ez}, {Tilve}, \& {Rodr{\'{\i}}guez-Llano}}]{CenarroOAJ}
{Cenarro}, A.~J., {Moles}, M., {Mar{\'{\i}}n-Franch}, A., {et~al.} 2014, in
  Society of Photo-Optical Instrumentation Engineers (SPIE) Conference Series,
  Vol. 9149, Society of Photo-Optical Instrumentation Engineers (SPIE)
  Conference Series, 1

\bibitem[{{Dawson} {et~al.}(2013){Dawson}, {Schlegel}, {Ahn}, {Anderson},
  {Aubourg}, {Bailey}, {Barkhouser}, {Bautista}, {Beifiori}, {Berlind},
  {Bhardwaj}, {Bizyaev}, {Blake}, {Blanton}, {Blomqvist}, {Bolton}, {Borde},
  {Bovy}, {Brandt}, {Brewington}, {Brinkmann}, {Brown}, {Brownstein}, {Bundy},
  {Busca}, {Carithers}, {Carnero}, {Carr}, {Chen}, {Comparat}, {Connolly},
  {Cope}, {Croft}, {Cuesta}, {da Costa}, {Davenport}, {Delubac}, {de Putter},
  {Dhital}, {Ealet}, {Ebelke}, {Eisenstein}, {Escoffier}, {Fan}, {Filiz Ak},
  {Finley}, {Font-Ribera}, {G{\'e}nova-Santos}, {Gunn}, {Guo}, {Haggard},
  {Hall}, {Hamilton}, {Harris}, {Harris}, {Ho}, {Hogg}, {Holder}, {Honscheid},
  {Huehnerhoff}, {Jordan}, {Jordan}, {Kauffmann}, {Kazin}, {Kirkby}, {Klaene},
  {Kneib}, {Le Goff}, {Lee}, {Long}, {Loomis}, {Lundgren}, {Lupton}, {Maia},
  {Makler}, {Malanushenko}, {Malanushenko}, {Mandelbaum}, {Manera}, {Maraston},
  {Margala}, {Masters}, {McBride}, {McDonald}, {McGreer}, {McMahon}, {Mena},
  {Miralda-Escud{\'e}}, {Montero-Dorta}, {Montesano}, {Muna}, {Myers},
  {Naugle}, {Nichol}, {Noterdaeme}, {Nuza}, {Olmstead}, {Oravetz}, {Oravetz},
  {Owen}, {Padmanabhan}, {Palanque-Delabrouille}, {Pan}, {Parejko},
  {P{\^a}ris}, {Percival}, {P{\'e}rez-Fournon}, {P{\'e}rez-R{\`a}fols},
  {Petitjean}, {Pfaffenberger}, {Pforr}, {Pieri}, {Prada}, {Price-Whelan},
  {Raddick}, {Rebolo}, {Rich}, {Richards}, {Rockosi}, {Roe}, {Ross}, {Ross},
  {Rossi}, {Rubi{\~n}o-Martin}, {Samushia}, {S{\'a}nchez}, {Sayres}, {Schmidt},
  {Schneider}, {Sc{\'o}ccola}, {Seo}, {Shelden}, {Sheldon}, {Shen}, {Shu},
  {Slosar}, {Smee}, {Snedden}, {Stauffer}, {Steele}, {Strauss}, {Streblyanska},
  {Suzuki}, {Swanson}, {Tal}, {Tanaka}, {Thomas}, {Tinker}, {Tojeiro},
  {Tremonti}, {Vargas Maga{\~n}a}, {Verde}, {Viel}, {Wake}, {Watson}, {Weaver},
  {Weinberg}, {Weiner}, {West}, {White}, {Wood-Vasey}, {Yeche}, {Zehavi},
  {Zhao}, \& {Zheng}}]{Dawson2013}
{Dawson}, K.~S., {Schlegel}, D.~J., {Ahn}, C.~P., {et~al.} 2013, \aj, 145, 10

\bibitem[{{D{\'{\i}}az-Garc{\'{\i}}a}
  {et~al.}(2015){D{\'{\i}}az-Garc{\'{\i}}a}, {Cenarro}, {L{\'o}pez-Sanjuan},
  {Ferreras}, {Varela}, {Viironen}, {Crist{\'o}bal-Hornillos}, {Moles},
  {Mar{\'{\i}}n-Franch}, {Arnalte-Mur}, {Ascaso}, {Cervi{\~n}o},
  {Gonz{\'a}lez-Delgado}, {M{\'a}rquez}, {Masegosa}, {Molino}, {Povi{\'c}},
  {Alfaro}, {Aparicio-Villegas}, {Ben{\'{\i}}tez}, {Broadhurst},
  {Cabrera-Ca{\~n}o}, {Castander}, {Fern{\'a}ndez-Soto}, {Husillos}, {Infante},
  {Aguerri}, {Mart{\'{\i}}nez}, {del Olmo}, {Perea}, {Prada}, \&
  {Quintana}}]{MUFFIT}
{D{\'{\i}}az-Garc{\'{\i}}a}, L.~A., {Cenarro}, A.~J., {L{\'o}pez-Sanjuan}, C.,
  {et~al.} 2015, \aap, in press [ArXiv:1505.07555]

\bibitem[{{Fitzpatrick}(1999)}]{Fitzpatrick1999}
{Fitzpatrick}, E.~L. 1999, \pasp, 111, 63

\bibitem[{{Gallego} {et~al.}(1995){Gallego}, {Zamorano}, {Aragon-Salamanca}, \&
  {Rego}}]{Gallego1995}
{Gallego}, J., {Zamorano}, J., {Aragon-Salamanca}, A., \& {Rego}, M. 1995,
  ApJL, 455, L1

\bibitem[{{Garn} \& {Best}(2010)}]{Garn2010}
{Garn}, T. \& {Best}, P.~N. 2010, \mnras, 409, 421

\bibitem[{{Geach} {et~al.}(2008){Geach}, {Smail}, {Best}, {Kurk}, {Casali},
  {Ivison}, \& {Coppin}}]{Geach2008}
{Geach}, J.~E., {Smail}, I., {Best}, P.~N., {et~al.} 2008, \mnras, 388, 1473

\bibitem[{{Gruel} {et~al.}(2012){Gruel}, {Moles}, {Varela},
  {Crist{\'o}bal-Hornillos}, {Ederoclite}, {Cenarro}, {Mar{\'{\i}}n-Franch},
  {Chueca}, {Yanes-D{\'{\i}}az}, {Rueda-Teruel}, {Rueda-Teruel}, {Luis-Simoes},
  {L{\'o}pez-Sainz}, \& {Hern{\'a}ndez Fuertes}}]{Gruel2012}
{Gruel}, N., {Moles}, M., {Varela}, J., {et~al.} 2012, in Society of
  Photo-Optical Instrumentation Engineers (SPIE) Conference Series, Vol. 8448,
  Society of Photo-Optical Instrumentation Engineers (SPIE) Conference Series,
  1

\bibitem[{Hunter(2007)}]{pylab}
Hunter, J.~D. 2007, Computing In Science \& Engineering, 9, 90

\bibitem[{{James} {et~al.}(2004){James}, {Shane}, {Beckman}, {Cardwell},
  {Collins}, {Etherton}, {de Jong}, {Fathi}, {Knapen}, {Peletier}, {Percival},
  {Pollacco}, {Seigar}, {Stedman}, \& {Steele}}]{James2004}
{James}, P.~A., {Shane}, N.~S., {Beckman}, J.~E., {et~al.} 2004, \aap, 414, 23

\bibitem[{{Kennicutt}(1992)}]{Kennicutt1992}
{Kennicutt}, Jr., R.~C. 1992, \apj, 388, 310

\bibitem[{{Kennicutt}(1998)}]{Kennicutt1998}
{Kennicutt}, Jr., R.~C. 1998, ARA`I\&'A, 36, 189

\bibitem[{{Kennicutt} \& {Kent}(1983)}]{Kennicutt1983}
{Kennicutt}, Jr., R.~C. \& {Kent}, S.~M. 1983, \aj, 88, 1094

\bibitem[{{Kewley} {et~al.}(2004){Kewley}, {Geller}, \& {Jansen}}]{Kewley2004}
{Kewley}, L.~J., {Geller}, M.~J., \& {Jansen}, R.~A. 2004, \aj, 127, 2002

\bibitem[{{Koyama} {et~al.}(2014){Koyama}, {Kodama}, {Tadaki}, {Hayashi},
  {Tanaka}, \& {Shimakawa}}]{Koyama2014}
{Koyama}, Y., {Kodama}, T., {Tadaki}, K.-i., {et~al.} 2014, ArXiv e-prints

\bibitem[{{Ly} {et~al.}(2011){Ly}, {Lee}, {Dale}, {Momcheva}, {Salim},
  {Staudaher}, {Moore}, \& {Finn}}]{Ly2011}
{Ly}, C., {Lee}, J.~C., {Dale}, D.~A., {et~al.} 2011, \apj, 726, 109

\bibitem[{{Ly} {et~al.}(2007){Ly}, {Malkan}, {Kashikawa}, {Shimasaku}, {Doi},
  {Nagao}, {Iye}, {Kodama}, {Morokuma}, \& {Motohara}}]{Ly2007}
{Ly}, C., {Malkan}, M.~A., {Kashikawa}, N., {et~al.} 2007, \apj, 657, 738

\bibitem[{{Marin-Franch} {et~al.}(2012){Marin-Franch}, {Taylor}, {Cepa},
  {Laporte}, {Cenarro}, {Chueca}, {Cristobal-Hornillos}, {Ederoclite}, {Gruel},
  {Hern{\'a}ndez-Fuertes}, {L{\'o}pez-Sainz}, {Luis-Simoes}, {Moles},
  {Rueda-Teruel}, {Rueda-Teruel}, {Varela}, {Yanes-D{\'{\i}}az}, {Benitez},
  {Dupke}, {Fern{\'a}ndez-Soto}, {Mendes de Oliveira}, {Sims}, {Sodr{\'e}}, \&
  {Toerne}}]{ToniT80Cam}
{Marin-Franch}, A., {Taylor}, K., {Cepa}, J., {et~al.} 2012, in Society of
  Photo-Optical Instrumentation Engineers (SPIE) Conference Series, Vol. 8446,
  Society of Photo-Optical Instrumentation Engineers (SPIE) Conference Series,
  6

\bibitem[{{Oke} \& {Gunn}(1983)}]{Oke83}
{Oke}, J.~B. \& {Gunn}, J.~E. 1983, \apj, 266, 713

\bibitem[{{Pascual} {et~al.}(2007){Pascual}, {Gallego}, \&
  {Zamorano}}]{Pascual2007}
{Pascual}, S., {Gallego}, J., \& {Zamorano}, J. 2007, \pasp, 119, 30

\bibitem[{{P{\'e}rez-Gonz{\'a}lez} {et~al.}(2008){P{\'e}rez-Gonz{\'a}lez},
  {Rieke}, {Villar}, {Barro}, {Blaylock}, {Egami}, {Gallego}, {Gil de Paz},
  {Pascual}, {Zamorano}, \& {Donley}}]{PerezGonzalez2008}
{P{\'e}rez-Gonz{\'a}lez}, P.~G., {Rieke}, G.~H., {Villar}, V., {et~al.} 2008,
  \apj, 675, 234

\bibitem[{{Salpeter}(1955)}]{Salpeter55}
{Salpeter}, E.~E. 1955, \apj, 121, 161

\bibitem[{{Salzer} {et~al.}(2000){Salzer}, {Gronwall}, {Lipovetsky}, {Kniazev},
  {Moody}, {Boroson}, {Thuan}, {Izotov}, {Herrero}, \& {Frattare}}]{Salzer2000}
{Salzer}, J.~J., {Gronwall}, C., {Lipovetsky}, V.~A., {et~al.} 2000, \aj, 120,
  80

\bibitem[{{Salzer} {et~al.}(2001){Salzer}, {Gronwall}, {Lipovetsky}, {Kniazev},
  {Moody}, {Boroson}, {Thuan}, {Izotov}, {Herrero}, \& {Frattare}}]{Salzer2001}
{Salzer}, J.~J., {Gronwall}, C., {Lipovetsky}, V.~A., {et~al.} 2001, \aj, 121,
  66

\bibitem[{{Sobral} {et~al.}(2009){Sobral}, {Best}, {Geach}, {Smail}, {Kurk},
  {Cirasuolo}, {Casali}, {Ivison}, {Coppin}, \& {Dalton}}]{Sobral2009}
{Sobral}, D., {Best}, P.~N., {Geach}, J.~E., {et~al.} 2009, \mnras, 398, 75

\bibitem[{{Sobral} {et~al.}(2012){Sobral}, {Best}, {Matsuda}, {Smail}, {Geach},
  \& {Cirasuolo}}]{Sobral2012}
{Sobral}, D., {Best}, P.~N., {Matsuda}, Y., {et~al.} 2012, \mnras, 420, 1926

\bibitem[{{Sobral} {et~al.}(2014){Sobral}, {Best}, {Smail}, {Mobasher},
  {Stott}, \& {Nisbet}}]{Sobral2014}
{Sobral}, D., {Best}, P.~N., {Smail}, I., {et~al.} 2014, \mnras, 437, 3516

\bibitem[{{Sobral} {et~al.}(2013){Sobral}, {Smail}, {Best}, {Geach}, {Matsuda},
  {Stott}, {Cirasuolo}, \& {Kurk}}]{Sobral2013}
{Sobral}, D., {Smail}, I., {Best}, P.~N., {et~al.} 2013, \mnras, 428, 1128

\bibitem[{{Takahashi} {et~al.}(2007){Takahashi}, {Shioya}, {Taniguchi},
  {Murayama}, {Ajiki}, {Sasaki}, {Koizumi}, {Nagao}, {Scoville}, {Mobasher},
  {Aussel}, {Capak}, {Carilli}, {Ellis}, {Garilli}, {Giavalisco}, {Guzzo},
  {Hasinger}, {Impey}, {Kitzbichler}, {Koekemoer}, {Le F{\`e}vre}, {Lilly},
  {Maccagni}, {Renzini}, {Rich}, {Sanders}, {Schinnerer}, {Scodeggio},
  {Shopbell}, {Smol{\v c}i{\'c}}, {Tribiano}, {Ideue}, \&
  {Mihara}}]{Takahashi2007}
{Takahashi}, M.~I., {Shioya}, Y., {Taniguchi}, Y., {et~al.} 2007, ApJS, 172,
  456

\bibitem[{{Thomas} {et~al.}(2013){Thomas}, {Steele}, {Maraston}, {Johansson},
  {Beifiori}, {Pforr}, {Str{\"o}mb{\"a}ck}, {Tremonti}, {Wake}, {Bizyaev},
  {Bolton}, {Brewington}, {Brownstein}, {Comparat}, {Kneib}, {Malanushenko},
  {Malanushenko}, {Oravetz}, {Pan}, {Parejko}, {Schneider}, {Shelden},
  {Simmons}, {Snedden}, {Tanaka}, {Weaver}, \& {Yan}}]{Thomas2013}
{Thomas}, D., {Steele}, O., {Maraston}, C., {et~al.} 2013, \mnras, 431, 1383

\bibitem[{{Villar} {et~al.}(2008){Villar}, {Gallego}, {P{\'e}rez-Gonz{\'a}lez},
  {Pascual}, {Noeske}, {Koo}, {Barro}, \& {Zamorano}}]{Villar2008}
{Villar}, V., {Gallego}, J., {P{\'e}rez-Gonz{\'a}lez}, P.~G., {et~al.} 2008,
  \apj, 677, 169

\bibitem[{{Zamorano} {et~al.}(1994){Zamorano}, {Rego}, {Gallego}, {Vitores},
  {Gonzalez-Riestra}, \& {Rodriguez-Caderot}}]{Zamorano1994}
{Zamorano}, J., {Rego}, M., {Gallego}, J.~G., {et~al.} 1994, \apjs, 95, 387

\end{thebibliography}
\bibliographystyle{aa}

\begin{appendix}
\section{3F method: Equations}\label{Apendice 3 filtros}
Here we develop the equations for the 3F method that was described in Sect.~\ref{metodologias 3 filtros}. The average flux $\overline{F}_{x}$ integrated any filter \emph{x} can be obtained, if the passband P properties are known, with the following expression:

\begin{equation}
  \label{convolucion}
  \overline{F}_{x}  =\frac{ \int F_{\lambda}P_{x}\left(\lambda\right)\lambda \mathrm{d}\lambda }{\int P_{x}\left(\lambda\right)\lambda \mathrm{d}\lambda}\:,
\end{equation}

\noindent where $P_{x}$ is the transmission of the passband \emph{x}, as a function of wavelength. 

Inside the \emph{r'} filter, there are two main contributions: the flux of the continuum $F_{r',\,\mathrm{cont}}$, and the flux of \Halpha + [\ion{N}{ii}], $F_{r',\,\mathrm{H}\alpha+\mathrm{[\ion{N}{ii}]}}$. Following the same reasoning, in the $F660$ filter there is $F_{F660,\,\mathrm{cont}}$ and $F_{F660,\,\mathrm{H}\alpha+\mathrm{[\ion{N}{ii}]}}$.\newline

\indent We approximate the continuum to a linear function

\begin{equation}\label{Continuo_r_apendice}
F_{\mathrm{cont}}\left(\lambda\right)=M\lambda+N,
\end{equation}

\noindent and for the emission, we assume that the flux of the three lines can be understood as one single, infinitely thin line centred at the \Halpha wavelength, which is described as a Dirac's delta function (the so-called infinite thin line approximation):

\begin{equation}\label{DeltaDiracApendice}
 F_{\mathrm{line}}\equiv F_{\mathrm{H}\alpha+\mathrm{[\ion{N}{ii}]}} \delta(\lambda-\lambda_{\mathrm{H}\alpha}).
\end{equation}

With this, the observed flux inside \emph{r'}:

\begin{equation}\label{r' en apendice}
\overline{F}_{r'}=\frac{\int F_{r',\,\mathrm{cont}}P_{r'}\left(\lambda\right)\lambda \mathrm{d}\lambda}{\int P_{r'}\left(\lambda\right)\lambda \mathrm{d}\lambda}+\frac{\int F_{\mathrm{line}}P_{r'}\left(\lambda\right)\lambda \mathrm{d}\lambda}{\int P_{r'}\left(\lambda\right)\lambda \mathrm{d}\lambda}.
\end{equation}

Plugging Eqs.~\ref{Continuo_r_apendice} and \ref{DeltaDiracApendice} into Eq.~\ref{r' en apendice}, and rewriting the integrals, we get

\begin{equation}\label{eq:r}
\overline{F}_{r'}=\frac{\int\left(M\lambda+N\right)P_{r'}\left(\lambda\right)\lambda \mathrm{d}\lambda}{\int P_{r'}\left(\lambda\right)\lambda \mathrm{d}\lambda}+\beta_{r'}F_{\mathrm{line}}=M\alpha_{r'}+\beta_{r'}F_{\mathrm{line}}+N\:,
\end{equation}

\noindent where $\alpha$ and $\beta$ can be defined at any passband \emph{x}, at any wavelength of interest $\lambda_{s}$, as

\begin{equation}
\alpha_{x}\equiv\frac{\int \lambda^{2}P_{x}\left(\lambda\right)\mathrm{d}\lambda}{\int P_{x}\left(\lambda\right)\lambda \mathrm{d}\lambda}\qquad\qquad\beta_{x}\equiv\frac{\lambda_{s}\mathrm{·}P_{x}\left(\lambda=\lambda_{s}\right)}{\int P_{x}\left(\lambda\right)\lambda \mathrm{d}\lambda}.
\end{equation}

In our case, $\lambda_{s}=\lambda_{\mathrm{H}\alpha}$. Following the same steps for the $F660$ and the \emph{i'} filters, we obtain

\begin{equation}
\overline{F}_{F660}=M\alpha_{F660}+\beta_{F660}F_{\mathrm{line}}+N\:,\label{660}
\end{equation}

\begin{equation}
\overline{F}_{i'}=M\alpha_{i'}+N\:.\label{eq:i}
\end{equation}

In Eq.~\ref{eq:i} the line contribution does not appear because the \emph{i'} filter does not cover it. Combining Eqs.~\ref{660} and \ref{eq:i} we obtain

\begin{equation}
M=\frac{\overline{F}_{F660}-\overline{F}_{i'}-\beta_{F660}F_{\mathrm{line}}}{\alpha_{F660}-\alpha_{i'}}.
\end{equation}

Plugging this into Eq.~\ref{eq:i}
\begin{equation}
N=\overline{F}_{i'}-\alpha_{i'}\left[\frac{\overline{F}_{F660}-\overline{F}_{i'}-\beta_{F660}F_{\mathrm{line}}}{\alpha_{F660}-\alpha_{i'}}\right]\:.
\end{equation}

With both parameters determined as a function of known values, we can go back to Eq.~\ref{eq:r} to obtain
\begin{equation*}
 F_{\mathrm{H}\alpha+\left[\mathrm{NII}\right]}=\frac{\left(\overline{F}_{r'}-\overline{F}_{i'}\right)-\left(\frac{\alpha_{r'}-\alpha_{i'}}
 {\alpha_{F660}-\alpha_{i'}}\right)\left(\overline{F}_{F660}-\overline{F}_{i'}\right)}{\beta_{F660}\left(\frac{\alpha_{i'}-\alpha_{r'}}{\alpha_{F660}-\alpha_{i'}}\right)+\beta_{r'}}\:,
\end{equation*}

\noindent which is Eq.~\ref{Ecflujo3filtro} from Sect.~\ref{metodologias 3 filtros}.

\end{appendix}

\end{document}